\documentclass[a4paper,11pt]{article}
\usepackage{latexsym}	
\usepackage{amsmath}
\usepackage{amsthm}

\usepackage{ amssymb }
\def\ba{\begin{eqnarray}}
\def\ea{\end{eqnarray}}

\def\ba{\begin{eqnarray}}
\def\ea{\end{eqnarray}}

\def\lb{\label}
\def\be{\begin{equation}}
\def\ee{\end{equation}}

\theoremstyle{plain}

\begin{document}
\title{The Maslov index and some applications to dispersion relations in curved space times}
\author{Juliana Osorio Morales \thanks{Departamento de Matem\'atica, FCEyN, Universidad de Buenos Aires, Buenos Aires, Argentina
juli.osorio@gmail.com.} and Osvaldo P. Santill\'an \thanks{Instituto de Matem\'atica Luis Santal\'o (IMAS), UBA CONICET, Buenos Aires, Argentina
firenzecita@hotmail.com and osantil@dm.uba.ar.}}
\date {}
\maketitle
\begin{abstract}
The aim of the present work is to generalize the results given in \cite{yo} to a generic situation for causal geodesics. It is argued that these results may be of interest for causality issues.
Recall that the presence of superluminal signals in a generic space time $(M, g_{\mu\nu})$ does not necessarily imply violations of the principle of causality \cite{hollowood1}-\cite{hollowood2}. In flat spaces,  global Lorenz invariance  leads to the conclusion that closed time like curves appear if these signals are present. In a curved space instead, there  is only local Poincare invariance, and the presence of closed causal curves may be avoided even in presence of a  superluminal mode, specially when terms violating the strong equivalence principle appear in the action. This implies that the standard analytic properties of the spectral components of these functions are therefore modified and, in particular, the refraction index $n(\omega)$ is not analytic in the upper complex $\omega$ plane. The emergence of this singularities may also take place for non superluminal signals, due to the breaking of global Lorenz invariance in a generic space time. In the present work, it is argued that the homotopy properties of the Maslov index \cite{maslov} are useful for studying how the singularities of $n(\omega)$ vary when moving along a geodesic congruence. In addition, several conclusions obtained in \cite{hollowood1}-\cite{hollowood2} are based on the Penrose limit along a null geodesic, and they are restricted to GR with matter satisfying strong energy conditions. The use of the Maslov index may allow a more intrinsic description of singularities, not relying on that limit, and a generalization of these results about non analiticity to  generic gravity models with general matter content. \end{abstract}
\section{Introduction}
The Maslov index is related to sympletic techniques and was originated by studying some quantum mechanical problems in the sixties \cite{maslov}. In the present work it will be argued that it may have several applications related to light propagation in curved space and, in particular, in studying causality issues.

The problem of causality violations is a subtle one. At first sight, the Einstein or some modified gravity equations may be solved by assuming that there are closed time like curves
in the space time, and by restricting then the matter fields to allow this behavior. The resulting solution will violate causality by construction. However, it is not clear that the resulting matter content will be physically reasonable. Loosely speaking, one of the faces of the problem is to understand if causality violation takes place
when reasonable matter fields  are present, or if it necessarily involves the presence of exotic matter fields.  In addition, it may be even difficult to  distinguish a reasonable matter field content from an exotic one on simple grounds.

In special relativity, the presence of superluminal particles induce causality violations. If there is a signal with velocity $w>c$ connecting two points at positions $x_1$ and $x_2$, the time required for this travel is $t_2-t_1=(x_2-x_1)/w$.
By a Lorentz transformation to a reference system with velocity $V$ the resulting time difference is $t'_2-t'_1=\gamma(t_2-t_1)(1-Vw)$. This implies that, if $V=1/w$, the interval is zero, which means that this observes sees a signal with infinite velocity.
For larger velocities, the observer sees an inversion of the time arrow. Thus, the resulting curve moves to the past. If this curve is combined with one going to the future and arriving at $x_2$ at some time before $t_2$, and furthermore the observer stays at this point $x_2$, then a closed time like curve has been constructed. This clearly contradicts causality.

 The above contradiction between superluminality and causality does not necessarily apply for curved space times. There are two main ingredients in this construction namely, the presence of a  superluminal mode and the existence of a global Lorentz transformation. For a curved space, global Lorentz invariance is lost and only local Poincare invariance remains. This may invalidate the reasoning of the previous paragraph \cite{shorecaus}-\cite{shoreaus3}.  

An attempt to insist in that superluminality should induce causality violations is to invoke the strong equivalence principle, which would play an analogous role of global Lorentz invariance. This is the statement that the laws of physics have the same form regardless the choice of the reference frame. However, this principle is not necessarily valid. In fact, there are suggestions that QED in curved space times involve corrections that apparently do violate this statement \cite{drummond}.  For instance, a coupling between the curvature and the Maxwell field of the form $F_{\mu\nu}R^{\mu\nu}$ induce a curvature dependence on the equations of motion, and thus the strong principle is not respected. It may be the case that superluminality is possible without violation of causality in a curved space time, specially when only the weak equivalence principle applies.

An important tool for studying causality are the characteristic surfaces of a generic wave solution, that is, the boundary between a perturbed and unperturbed zone for a propagating wave.
These regions move with a velocity equals to the high frequency limit of the phase velocity. This was illustrated for a one dimensional space dimension in generality already in 1937 \cite{landau}, and it is of interest
to review this short discussion here. Given a wave in one dimension, the phase and group velocities are given by 
  $$
  v_{ph}=\frac{\omega(k)}{k},\qquad v_g=\frac{d\omega}{dk}.
  $$
The wavefront velocity is defined as the velocity of the boundary between the perturbed and the unperturbed zones. In one dimension, any wave equation may be written in a first order formalism as
$$
a_{ij} \frac{d\phi_i}{dt}+b_{ij} \frac{d\phi_i}{dx}+c_{ij}\phi_j=0,\qquad \phi_i=\{u, \frac{\partial u}{\partial x}, \frac{\partial u}{\partial t}\}.
$$
A wave packet solution
$$
\phi_i=A_i e^{i\omega t-i k x},
$$
satisfies this equation if
\be\lb{deto}
[i\omega a_{ij}-i kb_{ij}+c_{ij}] A_j=0,\qquad \rightarrow \qquad \text{Det}[a_{ij}v_{ph}-b_{ij}-\frac{i}{k}c_{ij}] =0.
\ee
On the other hand, the wave front is a curve $(t, x(t))$ which separates the regions of $\phi_i=0$ and $\phi_i\neq 0$. At these points, the wave equation does not determine the solutions uniquely. Clearly $v_{wf}=x'(t)$ and a simple exercise of chain rule shows that the wave equation at a point $(t_0, x_0)$
in this front reduces to  
$$
[-a_{ij}v_{wf}+b_{ij}] \frac{\partial \phi_i}{\partial x}\bigg|_0+a_{ij} \frac{\partial \phi_i}{\partial t}\bigg|_0+c_{ij}\phi_j\bigg|_0=0.
$$
From here, it is seen that if 
$$
\text{Det}[a_{ij}v_{ph}-b_{ij}] \neq 0,
$$
the last equation will determine $(\phi_i)_x$ at the point $(t_0, x_0)$. This condition therefore does not represent a characteristic surface, since this region is composed by the points where the solution does not exist or it is not unique. Thus, 
the characteristic surface is defined by
$$
\text{Det}[a_{ij}v_{ph}-b_{ij}] = 0.
$$
By comparing this formula with (\ref{deto}) it is clear that
$$
v_{wf}=\lim_{k\to\infty}\frac{\omega}{k}.
$$
Thus, the wave front corresponds to the high frequency or the small wavelength limit of the phase velocity. These rays with very short wavelength are described by geometric optics. 

It is interesting to view causality from the perspective of response functions. In simple quantum systems, once a  Hamiltonian $H$ is perturbed by turning a source $s(x)$ the expectation value of a given operator $O$ at first order is perturbed as
$$
<\delta O>=\int \chi(t, t') s(t') dt',
$$
with $\chi(t, t')$ being the so called  linear response function. It is customary to assume that $\chi(t, t')=0$ when $t'<t$, as the perturbation cannot affect the past. In a flat space time, for time independent hamiltonians the effect of the perturbation is covariant under a time displacement, thus $\chi(t, t')=\chi(t-t')$. At the level of its Fourier components $\chi(\omega)$ these conditions lead to the 
well known property of analiticity in the upper complex plane $\omega$. In addition, it leads to the celebrated Kramers-Konig relations. In addition, in  a QFT in flat space, the analiticity property follows form the fact that the commutator of two operators $<0|[A(x)A(y)]|0>$ has to vanish outside the light cone. Usually, this commutator is also a function of $x-y$, due to global Lorenz transformations. This also leads to Lorenz invariance of the S matrix.

The considerations given above are not necessarily true in general gravity models with arbitrary matter content. In particular, if there are superluminal modes, then all the reasoning given above does not hold, as there is an argument between two different observers A and B about if the source was turned on before or after the perturbation. Thus, the analiticity property of $\chi(\omega)$ may be violated. In addition, the quantity $<0|[A(x)A(y)]|0>$  need not to vanish outside the light cone. But even when there are no superluminal signals, the standard logic leading to the usual dispersion relations has to be modified due to the fact that there is no global Lorenz invariance but only local Poincare invariance. In particular, there are quantities such as the vacuum refractive index $n(\omega, u)$ which do not necessarily satisfy the standard Kramers-Konig relations, even if $v_{wf}\leq1$. In fact, it is strongly suggested in \cite{hollowood1}-\cite{hollowood2},  that the standard dispersion relation at infinite frequency 
$$
n(\infty)=n(0)-\frac{2}{\pi}\int^\infty_0 \frac{\text{Im}n(\omega)d\omega}{\omega},
$$
is not true generically for a curved space time. If the medium is dispersive, that is, $\text{Im}n(\omega)\geq 0$, then $n(\infty)\leq n(0)$ which, together with this dispersion relation, implies that  $v_{wf}=v_f(\infty)\geq v_f(0)$. Thus, even if $v_f(0)$ is superluminal, this simply acts as a lower bound of light velocity.  But all these considerations are based on a Kramers-Konig type of relation which is doubtful in the curved context. The refraction index $n(\omega)$ needs not to be analytic at the upper half complex plane either. In fact, one of the interesting facts that the references \cite{hollowood1}-\cite{hollowood2} present is that the refraction index possess singularities in the upper half plane $\omega$ in QED in curved space times, even
without superluminality, in some plane wave background.

The appearance of singularities for $n(\omega)$ is related to the presence of conjugate points in null geodesics. These statements were obtained in some particular limit, which is known as a Penrose limit \cite{blau}-\cite{blau2}, which is adapted to the geodesic under study.
The purpose of this paper is to make a more intrinsic description of these matters, and to study the behaviour of the singularities when moving along a congruence of geodesics.

There is  second motivation studying the Maslov index in this context. The present authors already have worked out in \cite{yo} some results related the behavior of singularities in geodesics congruences with several applications in gravity theories, in particular, in the study of time delay or possible time advance. The advantage of the description in \cite{yo} in comparison with the present paper  is that it is less technical. However, we would like to take the chance to rediscover these results in a new formalism. Although more technical, the advantage is that the interpretation of these results is more obvious and there are some rich aspects related to the discretness of the set of conjugate points which are missing in \cite{yo}. In addition, the results of that reference are based on the use of the Raychaudhuri equation, and since this equation differ for time like and null geodesics, it can only compare both situations separately. The homotopy arguments given below does not have this limitation, and allows a comparison between causal curves in general. This generalization will be emphasized below, after the results are presented.

The present work is organized as follows. In section 2 we give a brief review about the role of conjugate points in the description of singularities of the refraction indices. The reader acquainted with this subject may skip this section. In section 3 the definitions of the focal and multiplicity index for a given causal geodesic is defined in certain detail, and a sympletic structure behind the Jacobi equation is clarified. This sympletic structure is compared with the results of the classical reference \cite{salamon} in section 5, and some homotopy properties related to the Jacobi problem are pointed out. By use of the results of that reference, the behavior of the singularities of the refraction index $n(\omega)$ when moving along a congruence of causal geodesics is clarified. It is shown that the appearence of caustic is generic, regardless of the matter content or the gravity theory under consideration. The physical significance of these results is analyzed in section 5. 

\section{The role of conjugate points in the singularities of $n(\omega)$}
It may be convenient at this point to illustrate the main points
which describe the relation between conjugate points and singularities of the refractive index. This description will be brief and schematic, the reader acquainted with all these ideas can skip directly to the following sections, and the reader searching for full details may consult the original references \cite{hollowood1}-\cite{hollowood2}.
\\

\textit{Polarization tensor in flat spaces}
\\

Before describing the refractive index in curved space times, it is convenient first
to characterize it in a flat space.

 Consider the propagator of a massive scalar particle with wave function $\phi(x)$, without electromagnetic fields turned on. This quantity can be expressed in the Schwinger representation as
$$
G(x, x')=<0|\hat{T}\phi(x)\phi(x')|0>=<x|\frac{1}{m^2-\square}|x'>=<x|\int_0^\infty e^{-T(m^2-\square)}|x'>.
$$
The contribution of the term with the D'Alambertian $\square$ is known from elementary path integral theory. By use of this formula the propagator can be expressed as 
\be\lb{world1}
G(x, x')=\int_0^\infty e^{-Tm^2} dT\int_{x(0)=x}^{x(T)=x'} Dx(t) e^{-\int_0^T\frac{\dot{x}^2}{4}dt}.
\ee
This is the world line representation of the propagator \cite{feynman}-\cite{schwinger}, \cite{strassler}-\cite{schubert}.
By decomposing the trajectory as the sum of the classical one plus a fluctuation $x(t)=x_{cl}(t)+q(t)$, with $q(0)=q(T)=0$, the Green function becomes
$$
G(x, x')=\int_0^\infty e^{-Tm^2}  dT e^{-\frac{(x-x')^2}{4T}}\int_{q(0)=q(T)=0} Dq(t) e^{-\int_0^T\frac{\dot{q}^2}{4}dt}.
$$
The integral related to the fluctuations $q$ is given by $(4\pi T)^{-D/2}$, with $D$ the space time dimensions. In a curved space however, this quantity
presents a more complicated behaviour and will be responsible for the singularities of the refractive index, as it will be shown below.

When background electromagnetic fields are turned on and the scalar field has charge $e$, the quantity given above is generalized to \cite{schubert}
$$
G(x, x', A)=<x|\frac{1}{m^2-(\partial +i e A)^2}|x'>=\int_0^\infty e^{-Tm^2} dT\int_{x(0)=x}^{x(T)=x'} Dx(t) e^{-\int_0^T[\frac{\dot{x}^2}{4}+ie A\cdot x]dt},
$$
with $A^\mu$ the corresponding gauge potential. If the exponentials 
containing $A^\mu$ are expanded in Taylor series, then this quantity leads to the standard Feynmann diagrams for bosons interacting with the external background field $A^\mu$, . 

For studying the corrections to the Maxwell equation in vacuum due to virtual pair creation and annihilation, a crucial role is played by the so called effective action. Recall that the partition function for scalar QED is given by
$$
Z(j)=\int DA D\phi \;e^{\int [-\frac{i}{4} F_{\mu\nu}F^{\mu\nu}+i D_\mu \phi D^\mu \phi+V(\phi)+A^\mu j_\mu]d^4x},\qquad D_\mu \phi=\partial_\mu \phi+ ie A_\mu\phi,
$$
where $\phi$ is the scalar field, $e$ it charge, $V(\phi)$ its energy density and $j_\mu$ is a source turned on. The effective action $\Gamma_{eff}(A)$ is  defined by integrating out the scalar degrees of freedom
$$
e^{i\Gamma_{eff}(A)}=\int D\phi \;e^{i\int [D_\mu \phi D^\mu \phi+V(\phi)+A^\mu j_\mu]d^4x}.
$$
In these terms the partition becomes a function of the gauge potentials only
$$
Z(j)=\int DA\;e^{\int [-\frac{i}{4} F_{\mu\nu}F^{\mu\nu}+A^\mu j_\mu]d^4x+i\Gamma_{eff}(A)},
$$
and consequently the Maxwell equations are modified by the addition of this effective terms.
The expansion of the effective action around a given classical field $A^0$ is given by
$$
\Gamma_{eff}(A)=\Gamma^0_{eff}(A_0)+\int d^4x \Gamma^2_{eff,\mu\nu}(A_0) \delta A^\mu \delta A^\nu+\text{higher orders}.
$$
The discussion given above shows that, at low order, there will be corrections to the Maxwell equations  of the form
$$
\delta S_{qc}=\int d^4x d^4x' A^\mu(x) \Pi_{\mu\nu}(x,x') A^\nu(x').
$$
The so called polarization tensor $\Pi_{\mu\nu}$ introduced in the last formula contains the loop Feynmann diagrams related to vacuum pair creation and annihilation. 

In the world line formalism, the effective action giving rise to the diagrams contributing to $\Pi_{\mu\nu}(x,x')$ is given in terms of a path integral \cite{schubert}
$$
\Gamma(A)=\int_0^\infty e^{-Tm^2} \frac{dT}{T}\int_{x(0)=x(T)} Dx(t) e^{-\int_0^T[\frac{\dot{x}^2}{4}+ie A\cdot x]dt},
$$
where, unlike for $G(A, x, x')$, the initial and final points are identified, $x(0)=x(T)$. The initial point of the trajectory along the corresponding circle is unspecified. The additional 
factor $1/T$, which is not present in the definition of $G(A, x, x')$  takes care about this ambiguity, as is proportional to the circle length.
In addition, the absence of initial and final points suggest that the diagrams that this object generate are 1-particle loop irreducible for photons.
In fact, it does. The polarization tensor $\Pi_{\mu\nu}$ is calculated in terms of $\Gamma(A)$  by making the expansion of the gauge potential
$$
A_i^\mu=A(x)\epsilon_i^\mu(x) e^{i k_i x},
$$
with $A(x)$ an amplitude, and by Taylor expanding the corresponding exponential in $\Gamma(A)$. At second order, the result is identified with the 1-loop polarization, namely
$$
\Pi^{\text{1-loop}}_{\mu\nu}(x,x')=\frac{\alpha}{4\pi}\int_0^\infty \frac{dT}{T^3}e^{-Tm^2}\int_0^T d\tau \int_{x(0)=x(T)} Dx(t) e^{-\int_0^T\frac{\dot{x}^2}{4}dt} V_{\omega, \epsilon_i}[x(\tau)] V_{\omega, \epsilon_j}[x(0)],
$$
where the additional factor $T^{-2}$ comes from the integration of the zero mode $x_0^\mu$. Note that the result does not depend on $x_0^\mu$ on a flat space due to translational symmetry, but it may depend on it in a curved space.
The vertex operators  in the last expression come from the expanded gauge potential, and are given by
$$
V_{\omega, \epsilon}[x]=A(x)\dot{x}\cdot\epsilon_i(x) e^{i k_i x}.
$$
As discussed above, the limit of geometric optics $\omega>>1$, is the one which describe the velocity of the wavefronts. In this limit  the gauge potential takes the form
$$
A^\mu\sim \bigg(A(x) \epsilon^\mu(x)+\frac{B^\mu(x)}{\omega}+..\bigg) e^{i\omega\Theta(x)},
$$
where $\Theta(x)$ is a rapidly oscillating phase and $A(x)\epsilon^\mu(x)$ a
 slowly changing function. Then the polarization tensor may be expressed as
 $$
\Pi^{\text{1-loop}}_{\mu\nu}=\frac{\alpha}{4\pi}\int_0^\infty \frac{dT}{T^3}\int_0^T d\tau <A(x(\tau))\epsilon_i(x(\tau))\cdot x(\tau)\; A(x(0))\epsilon_i(x(0))\cdot x(0)>_m
$$
where the modified average $<>_m$ is now given in terms of the modified action
$$
S_m=\int_0^T \bigg[\frac{\dot{x}}{4}^2-m^2\bigg] d\tau-\omega \Theta[x(\tau)]+\omega \Theta[x(0)].
$$
The insertion of the phase dependent terms modify the action, and the average  $<>_m$ can be studied by considering the trajectories corresponding to this action, which
will differ from classical trajectories due to the insertion terms proportional to $\Theta$.
\\

\textit{Refraction index in curved space time}
\\

For curved spaces, the contribution of the fluctuations $q(T)$ considered above is more involved, and is directly related to the presence of conjugated points. 
For simplicity, consider first the Green function for a massless scalar field. The generalization of (\ref{world1}) in a curved setting is
\be\lb{world2}
G(x, x')=\int_{x(0)=x}^{x(T)=x'} D[x^\mu(t)] e^{-S[x^\mu(t)]}, \qquad S[x^\mu(t)]=\frac{1}{4}\int_0^T g_{\alpha\beta}(x^\mu(t))\frac{dx^\alpha}{dt}\frac{dx^\beta}{dt} dt.
\ee
 The classical trajectory corresponding to this action, if there are no conjugate points, are geodesics connecting $x$ and $x'$, which will be denoted as $z^\mu(t)$.  A non classical trajectory $x^\mu(t)$ will be decomposed as $x^\mu=z^\mu+y^\mu$ , with the fluctuation $y^\mu(\lambda, t)$ represented as a geodesic segment connecting the points at equal $t$ of these trajectories. Here $\lambda$ is an affine parameter, not to be confused with the evolution one. The derivative 
$$
q(t)=-\frac{dy^\mu}{d\lambda}\bigg|_{\lambda=0},
$$
is such that $q^\mu(0)=q^\mu(T)=0$. The expansion of the geodesic action up to second order in $q$ is
$$
S[x^\mu(t)]=\frac{\sigma(x, x')}{2T}+\frac{1}{4}\int_{0}^T[\dot{q}^\alpha\dot{q}_\alpha-R_{\mu\nu\alpha\beta}\dot{q}^\mu \dot{q}^\alpha u^\nu u^\beta] dt.
$$
Here $\sigma(x, x')$ is the action $S$ in (\ref{world2}) evaluated along the geodesic, and represents the geodesic distance between the initial and final points. This quantity is known as the Synge bitensor. On the other hand $u^\nu$
is the unit vector along $z^\mu(t)$. By going to an integration in $D[q^\mu(t)]$
the propagator becomes in this WKB approximation
$$
G(x, x')= e^{\frac{-\sigma(x, x')}{2s}}\int D[q^\mu(t)] e^{-\frac{1}{4}\int_{0}^s[\dot{q}^\alpha\dot{q}_\alpha-R_{\mu\nu\alpha\beta}\dot{q}^\mu \dot{q}^\alpha u^\nu u^\beta]}dt,
$$
with $q^\mu(0)=q^\mu(s)=0$. The last is the generalization of the flat space fluctuation factor $\int Dq(t) e^{-\int_0^T\frac{\dot{q}^2}{4}dt}$ with periodic conditions, to a curved space.
By  use of Fermi coordinates, it is found that \cite{zannias}-\cite{schulman}
\be\lb{gree}
G(x, x')= \frac{1}{\sqrt{{\text{Det}J^\alpha_\beta}}}e^{\frac{-\sigma(x, x')}{2T}}.
\ee
Here the quantity $J^\alpha_\beta$ is the solution of the system of equations
$$
\frac{D J^\alpha_\beta}{dt^2}=-R^\alpha_{\beta\gamma\delta} u^\beta u^\delta J^\gamma_\beta,\qquad J^\alpha_\beta(0)=0,\qquad \frac{DJ^\alpha_\beta(0)}{dt}=\delta_\beta^\alpha,
$$
with $D/dt$ denoting the derivative along the geodesic $z^\mu$.
The quantity $J^\alpha_\beta$ is crucial for describing conjugate points. This is seen by the fact that the vector $v^\alpha(\tau)=J^\alpha_\beta \dot{v}^\beta(0)$
is a solution of the Jacobi equation
$$
\frac{D^2 v^\alpha}{dt^2}=-R^\alpha_{\beta\gamma\delta} u^\beta u^\delta v^\gamma,\qquad v^\alpha(0)=0,\qquad \frac{dv^\alpha(0)}{dt}=I^\alpha.
$$
It is known that conjugate points correspond to solutions for which $v^\alpha(t)=0$ for some $t$ value. Such points will appear if and only if the determinant of $J^\alpha_\beta$
vanish. In other words, this matrix will have a zero eigenvalue. When conjugate points appear, the Green function becomes singular due to the corresponding zero of the denominator in (\ref{gree}).

By completeness, it should be mentioned \cite{zannias} that this quantity is related to  Van Vleck De Witt determinant $\Delta(x, x')$, in such a way that the propagator may be written as
$$
G(x, x')= \frac{1}{(4\pi T)^{\frac{D}{2}}}e^{\frac{-\sigma(x, x')}{2T}} \sqrt{\Delta(x, x')}.
$$
Here the Van Vleck De Witt determinant is given by
$$
\Delta(x, x')=\frac{1}{\sqrt{g(x) g(x')}} \text{Det}\bigg(-\frac{\partial^2 \sigma}{\partial x^\mu \partial x'^\nu}\bigg).
$$
For a massive particle it may be correct to write the full expression as
$$
G(x, x')= \sqrt{\Delta(x, x')}\int_0^\infty \Omega(T, x, x') e^{\frac{-i m^2 s-\sigma(x, x')}{2T}} \frac{dT}{(4\pi T)^{\frac{D}{2}}}.
$$
In all these formulas $\Omega(T, x, x')$ encode higher order curvature terms \cite{hollowood1}. The Van Vleck De Witt determinant is singular at the conjugate points.

Analogous considerations follow for the effective action $\Gamma(A)$ in curved spaces. In fact, the reason for which conjugate points
give rise to singularities can be visualized by studying the polarization tensor 
 $$
\Pi^{\text{1-loop}}_{\mu\nu}=\frac{\alpha}{4\pi}\int_0^\infty \frac{dT}{T^3}\int_0^T d\tau <A(x(\tau))\epsilon_i(x(\tau))\cdot x(\tau)\; A(x(0))\epsilon_i(x(0))\cdot x(0)>_m,
$$
where the modified average $<>_m$ is is given in terms of the modified action \cite{hollowood1}-\cite{hollowood2}
$$
S_m=\int_0^T \bigg[\frac{g_{\mu\nu}\dot{x}^\mu \dot{x}^\nu}{2}-m^2\bigg] d\tau-\omega \Theta[x(\tau)]+\omega \Theta[x(0)].
$$
The equation of motion corresponding to this action are given by geodesics except at some delta singularities induced by the phase terms $\Theta(x)$.
As is well known, the solution is unique if there are no conjugate points in between. Instead, if conjugate points appear in the middle of the two singularities, then 
a continuous set of classical solutions may appear, resulting in zero modes for some $\tau$ values. These modes induce singularities in the polarization tensor.

The propagation of photons, at quantum level, is determined by the vacuum polarization term given by
\be\lb{pove}
\nabla_\mu F^\mu_{\nu}=\int A^\nu(x') \Pi_{\mu\nu}(x, x')\sqrt{g(x')} dx'.
\ee
On the other hand, the eikonal level the field is given by
$$
A_i^\mu(x)=A(x)\epsilon_i^\mu(x) e^{i\Theta(x)},
$$
with $\Theta(x)$ a rapidly varying phase, must be corrected in this approximation. The quantum one loop corrections induce a polarization dependent deviation of the phase, and the field becomes
$$
A_i^\mu(x)=A(x)\epsilon_i^\mu(x) e^{i(\Theta(x)-\theta_{ij})},
$$
for which the field equations get corrected as
\be\lb{pove2}
\nabla_\nu F_{(i)\mu}^\nu=2\omega^2\frac{\partial \theta_{ij}}{du}\epsilon^\nu_{j} e^{-i\omega v},
\ee
where the coordinate $v(x)$ is defined by the relation $\Theta(x)=\omega v$.
The refraction index is then given by 
$$
n_{ij}(u, \omega)=1+2\frac{\partial \theta_{ij}}{du}.
$$
Comparison between (\ref{pove}) and (\ref{pove2}) allows to find an expression for the refraction index in terms of the described Green functions.
The result is
$$
n_{ij}(x,\omega)=\delta_{ij}-\frac{2}{\omega^2 A(x)}\epsilon^\nu_{i}(x)e^{-i\omega v}\int \sqrt{g(x')}\Pi_{\mu\nu}(x,x')A(x')\epsilon^\mu_j(x') e^{i\omega v'} dx'.
$$
These formulas were heavily employed in references \cite{hollowood1}-\cite{hollowood2}. This discussion shows that the presence of conjugate points
give rise to singularities on the polarization tensors.

\section{An intuitive but formal picture about the Maslov index and its uses}

As discussed above, the presence of conjugate points implies that the polarization tensor $\Pi_{\mu\nu}(x, x')$ becomes singular. It is likely that such
singularities translate to ones for $n_{ij}(x,\omega)$. However, there exist primitives with singular integrands which are perfectly regular, so a more careful analysis
should be performed. The authors \cite{hollowood1}-\cite{hollowood2} study these singularities, in the Penrose limit for a given geodesic.
The Penrose limit assigns to every space time ($M$, $g_{\mu\nu}$) and to a given null geodesic $\gamma$ a limiting plane wave metric. 
First, given the null geodesic $\gamma$, a possible choice of coordinates may be $v$ and the geodesic parameter $u$.
The metric in these adapted to $\gamma$ coordinates reads
$$
g=2 du dv+a(u, v, y^k)dv^2+2 b_i(u, v, y^k)dv dy^i+g_{ij}(u, v, y^k)dy^i dy^j.
$$
These coordinates always exist locally, see for example \cite{blau}-\cite{blau2}.  The only special feature about this choice is that the term proportional to $du dv$ is a simple constant. This metric corresponds on taking a null geodesic $\gamma$ parameterized by the affine parameter $u$, and embedding it into a congruence of geodesics parameterized by $v$. The choice of the spatial coordinates $y^k$ is not relevant in the following. By performing the scaling of coordinates 
$$
(U, V, Y^k)=(u, \lambda^2 v, \lambda y^k),
$$
and taking the limit  $\lambda\to 0$ in the expression $g_p=\lambda^{-2}g$ the resulting metric becomes
$$
g_p=dU dV+g_{ij}(U)dY^i dY^j,
$$
with $g_{ij}(U)=g_{ij}(U, 0, 0)$. There exists a change of coordinates which brings the last expression to the form
of a gravitational wave
$$
g_p=du d\eta +h_{ij}(u) x^i x^j d\eta^2+\delta_{ij}dx^i dx^j.
$$
Here the profile $h_{ij}(u)=R_{uiuj}|_\gamma$ is related to the curvature tensor of the original space time evaluated at the null geodesic $\gamma$ \cite{shorepenrose}. By assuming that
there are conjugate points for the given geodesic, for some particular profiles, these authors are able to show that singularities on $n_{ij}$ do appear at the upper $\omega$ complex plane.
These singularities do not necessarily appear only for superluminal modes.

The arguments of those references are, in the authors opinion, pretty solid. But there are some issues that we would like to improve.  The results are only related to plane wave profiles, which are designed to specially admit caustics,
 that is, points where the geodesics of the congruence intersect. The presence of such caustics generate a full region composed of conjugate points. This motivates the use of the Brinkman coordinates, which are regular at this region.
However, it may be argued that these caustics are just an artifact of the Penrose limit, which not necessarily lift to the exact metric before this limit was taken. The aim of the present work is to show
that, regardless the gravity model or the matter content, these caustics are for real, if the geodesic under study does admit a conjugate point. In other words, the results obtained in those references, even approximated,
capture the real physics of what happens around the chosen geodesic. To show such result requires some technical tools. It will be shown that the Maslov index techniques introduced in \cite{maslov} and further developed
in references \cite{robbin1}-\cite{leray}
may be helpful for showing the presence of such caustics. 

 For further applications related to quantum mechanics, variational problems or to spectral theory of operators the reader may consult the classical works \cite{robbin1}-\cite{leray}.

\subsection{Some indices associated to conjugate points in causal geodesics}\label{indices associated}

\emph{The case for timelike geodesics}
\\

Even at cost of reviewing some elementary topics, it is convenient to discuss some details that may be confusing when reading \emph{correct} mathematical and physical literature
about geodesic deviation. In some excellent physics textbooks such as \cite{wald}-\cite{poisson} there is an approach about the topic that relies on a congruence of geodesics. In this approach, one considers  a time like geodesic $\gamma$ in an $n$-dimensional space-time ($M$, $g$)  with $n-1$ spatial dimensions. This space time is assumed to be time orientable, in the sense that there exists  a nowhere vanishing future directed time vector $t^\mu$ in $M$.
A system of local coordinates $x^\alpha$ may be chosen, for which the geodesic $\gamma$ under study may be embedded  into a particular congruence of geodesics
$x^\alpha(t,s)$, with $s$ a one dimensional parameter distinguishing the different curves in the congruence. In particular, $\gamma$ is identified with $x^\alpha(t,0)$. The tangent vector to these curves
$$
u^\alpha=\frac{\partial x^\alpha}{\partial t},
$$
satisfies the geodesic equation $u^\beta\nabla_\beta u^\alpha=0$. 

On the other hand, for fixed $t$ the vector $x^\alpha(t, s)$ describe another set of curves, not necessarily geodesics.
The tangent to these curves is given by the so called deviation vector
$$
v^\alpha=\frac{\partial x^\alpha}{\partial s}.
$$
The property in flat spaces that $\partial_t \partial_s x^\alpha=\partial_s\partial_t x^\alpha$ or, what is the same, $\partial_t v^\alpha=\partial_s u^\alpha$, is translated to a curved space into
 the identity ${\cal L}_u v=[u,v]=0$ with ${\cal L}_X$ denoting the standard Lie derivative along the vector field $X$. This is equivalent to the condition
 \be\lb{ido}u^\alpha \nabla_\alpha v^\beta=v^\alpha \nabla_\alpha u^\alpha. \ee This identity, together with the geodesic equation, implies that the quantity $v^\alpha u_\alpha$ is constant along the geodesic \cite{poisson}-\cite{wald}. In fact, a simple calculation shows that
$$
\frac{d}{dt}(v^\alpha u_\alpha)= u^\alpha u^\beta \nabla_\beta v^\alpha=u^\alpha v^\beta\nabla_\beta u^\alpha=0,
$$
where in the last step (\ref{ido}) was employed. It is concluded that, if $v^\alpha u_\alpha=0$ is imposed at the initial time $t_0$ then the deviation vector $v^\alpha$ may be chosen orthogonal to the tangent $u^\alpha$ of the geodesic. The acceleration 
$$
a^\alpha=\frac{D^2 v^\alpha}{dt^2}\qquad \frac{D}{dt}=u^\alpha \nabla_\alpha,
$$
can be worked out by curvature identities, by (\ref{ido}), and by the fact that the length of $u^\alpha$ is conserved along $\gamma$.
The calculation throws
$$
\frac{D^2 v^\alpha}{dt^2}=u^\gamma \nabla_\gamma u^\beta \nabla_\beta v^\alpha=u^\gamma \nabla_\gamma v^\beta \nabla_\beta u^\alpha
=u^\gamma v^\beta \nabla_\gamma \nabla_\beta u^\alpha+(u^\gamma \nabla_\gamma v^\beta) \nabla_\beta u^\alpha
$$
$$
=u^\gamma v^\beta (\nabla_\beta \nabla_\gamma u^\alpha-R^\alpha_{\alpha\delta\beta}u^\delta)+(v^\beta \nabla_\beta u^\gamma) \nabla_\gamma u^\alpha
$$
The sum of the first and the last term are proportional to the geodesic term $u^\alpha \nabla_\alpha u^\beta$, which clearly vanishes. 
In this manner, the well known Jacobi equation \cite{poisson}-\cite{wald}
 \be\lb{jaco2}
\frac{D^2 v^\alpha}{dt^2}=-R^\alpha_{\beta\sigma\delta} u^\beta v^\sigma u^\delta,\qquad \frac{D}{dt}=u^\alpha \nabla_\alpha,
\ee
is obtained. The fields $v^\alpha$ are denominated Jacobi fields. 

Take the geodesic under study $\gamma$ at some time $t=t_0$, corresponding to a point $p$ in $M$. Solve equation (\ref{jaco}) together with the initial conditions
\be\lb{inito}
 v^\alpha(t_0)=0,\qquad \dot{v}^\alpha(t_0)=I^\alpha,
\ee
where $I^\alpha$ is an arbitrary vector field. If, for some choice of $I^\alpha$, there is a point $q$ corresponding to a time parameter $t_1$ such $v^\alpha(t_1)=0$, then $q$ is known as a conjugated point to $p$.

The approach given above is transparent and intuitive. It suggest that conjugate point arises when gravity is such that geodesics pointing to different directions are focused at a given point. However, there is a subtle detail in all this derivation. In this deduction, the property (\ref{ido}) has been employed. As remarked above, this imply that $v^\alpha u_\alpha$ is constant along $\gamma$. This fact, together with the initial conditions (\ref{inito}) would imply that $$v^\alpha u_\alpha=0,$$
along the geodesic. As this constancy was heavily employed in finding the Jacobi equation it is then attractive to state that every solution $v^\alpha$ of (\ref{jaco2}) with the conditions (\ref{inito}) should be orthogonal to $u^\alpha$. However, this affirmation is \emph{false}, even taking into account the last formula. In fact, the vector $v^\alpha=(t-t_0)u^\alpha$ is a solution of (\ref{jaco2}) with these initial conditions, and clearly $v^\alpha u_\alpha$ is not constant for this solution.

The subtlety described above does not indicate that the approach of \cite{poisson}-\cite{wald} is wrong. In fact, one may use a series of hypothesis for finding an equation and then realize that the spectra of solutions is  wider than expected.
For the purposes of the present work however, this point is to be remarked. The reason is that some mathematical indices will be defined below, which are based on the spectrum of solutions of the Jacobi equation, regardless this is applied for studying geodesic congruences or not. More mathematical oriented references such as \cite{hawking}-\cite{penrose} or even the cited textbooks \cite{poisson}-\cite{wald} in other chapters deduce the Jacobi equation in terms of the second variation of the geodesic length functional, and show that the presence of conjugate point spoil extremal properties of time like geodesics. In this approach, the orthogonality property $u_\alpha v^\alpha=0$ does not play any significant role.

Note that the solution $v^\alpha=(t-t_0)u^\alpha$ does not correspond to a conjugate point since it will never vanish for $t\neq t_0$, thus it makes sense to restrict the attention to the solutions orthogonal to $u^\alpha$, as the authors of \cite{poisson}-\cite{wald} do. However this restriction should be done explicitly, the equation (\ref{jaco2}) with (\ref{inito}) alone does not warrant orthogonality with $u^\alpha$.

Some simple versions of the equation (\ref{jaco2}) are in order. At the initial point $p$ corresponding to the time parameter $t_0$, an orthonormal basis $e_a(t_0, s)$ of  $TM_{p(s)}$, that is, a basis such that at $p$ the metric becomes diagonal
\be\lb{orto2}
g=\eta_{ab} e^a \otimes e^b,
\ee
may be constructed, which in addition satisfy $e_0(t_0, s)=u^\alpha(t_0, s)\partial_\alpha$. If this basis is parallel transported along the geodesic $\gamma$, that is, $u^\alpha \nabla_\alpha e_a=0$ then 
\be\lb{orto2}
g=\eta_{ab}e^a\otimes e^b,\qquad e_0(t, s)=u^\alpha(t,s )\partial_\alpha,
\ee
with $e^a$ the dual basis to $e_a$. This formula is valid for any $t$ in a neighborhood of $t_0$. 
Note that the transverse $n-1$ dimensional metric
\be\lb{espasial}
h=\delta_{ab} e^a\otimes e^b,\qquad a,b=1,..,n-1,
\ee
is spatial. The Jacobi equation (\ref{jaco2}) in this basis is simplified as
 \be\lb{jacoss}
\frac{d^2 v^a}{dt^2}=-R^a_{0c0}  v^c.
\ee
This expression will make some of the calculations below easier.

 An important concept is the multiplicity of a conjugate point $q$ to $p$, defined as follows. By  varying over the possible choices of $I^\alpha$ in (\ref{init}) the space of non trivial Jacobi fields $J[\gamma]$ on the geodesic is found. At some point $q$ with parameter $t_1$, there is a set of $k$ Jacobi fields $v_l$  with $l=1,..,k$ that vanish, i.e,
 $v_l^\alpha(t_1)=0$ with $\alpha=1,..,n$. The multiplicity of the conjugate point $q$ is defined as
\be\lb{multi}
\text{mult}(q)=k.
\ee
As the solution $v^\alpha=(t-t_0)u^\alpha$ never represents a conjugate point, it is clear that $\text{mult}(q)\leq n-1$.

Given the characterization above of the time like geodesics $\gamma:[a,b]\to M$ in a generic space time ($M$, $g$) and its conjugate points, two useful indices may be introduced. One of them is the geometric index of $\gamma$, which is given in terms of the multiplicity (\ref{multi}) as
\be\lb{igeom}
i_{geom}(\gamma)=\sum_{q\in [a,b]} \text{mult} (q).
\ee
This quantity is naturally related to properties of  conjugate points along geodesics. But there is another index defined in the literature, known as the focal index $i_{focal}(\gamma)$, which will be useful in the present exposition as well.

The definition of the focal index $i_{focal}(\gamma)$ is a bit more involved than (\ref{igeom}). In order to define it, consider the Jacobi problem with initial conditioins:
 \begin{align*}
     &\frac{d^2 v^a}{dt^2}=-R^a_{0c0}  v^c\\
     & v^\alpha(t_0)=0,\qquad \dot{v}^\alpha(t_0)=I^\alpha,
 \end{align*}

 Given the initial point $p$ corresponding to $t_0$, then  at a time $t_1$ a conjugate point $q$ of multiplicity $k$ appears if there 
are $k$ choices of linearly independent Jacobi fields $v_i(t)$ with $i=1,..,k$ such that $v_i(t_1)=0$. Consequently, in this situation,  there are $n-k$ choices of Jacobi fields  $v_l(t)$ with $l=k+1,.., n$ for which
 $v_i(t_1)\neq 0$. In particular,  the solution $v^\alpha=(t-t_0)u^\alpha$. These non vanishing vectors $v_i(t_1)$ generate the space of non trivial Jacobi vectors at $t_1$, denoted as $\mathbb{J}[t_1]$. 
Denote orthogonal complement of $\mathbb{J}[t_1]$ in $T_q M$ as $\mathbb{J}[t_1]^\perp$. Note that this complement never contains the direction defined by $u^\alpha$, as the corresponding fields never vanish. An arbitrary basis $b_a$ at $T_qM$ can be chosen for this complement, with $a=1,..,k$. Construct then the matrix $\widetilde{g}_{ab}=g(b_a, b_b)$, which is the restriction of $g$ on the space $\mathbb{J}[t_1]^{\perp}$  with respect the basis $b_a$. Denote by $n_\pm(\widetilde{g})$ the maximal dimension of a subspace $V\in \mathbb{J}[t_1]^\perp$ such that the matrix $\widetilde{g}_{ab}$ is positive or negative definite respectively, and $n_0(\widetilde{g})$
the dimension of its kernel. It is clear that $$n_+(\widetilde{g})+n_-(\widetilde{g})+n_0(\widetilde{g})=k,$$in the present case. The signature of $\widetilde{g}_{ab}$ is defined as
$$
\text{signature}(g|_{\mathbb{J}[t_1]^{\perp}})=n_+(\widetilde{g})-n_-(\widetilde{g}).
$$
The signature defined above is of course independent on the choice of the basis $b_a$.
In these terms, the focal index of $\gamma$ is given by
\be\lb{ifocal}
i_{focal}(\gamma)=\sum_{t\in [a,b]} \text{signature} (g|_{\mathbb{J}[t]^{\perp}}).
\ee
At first sight, there is no reason for the two indices (\ref{igeom}) and (\ref{ifocal}) to be identified. If the multiplicity of a point $q$ is $k$, then the dimension of $g|_{\mathbb{J}[t_1]^{\perp}}$ is also $k$. However, its signature can differ from $k$. This means that in a generic problem of semi-riemannian geometry with arbitrary signature, both indices may have different values. In addition, for a semi-riemannian geometry, it may be possible to reach a region of which $g|_{\mathbb{J}[t_1]^{\perp}}$ is degenerate at a conjugate point, even if the full metric is regular there. For example, when studying a space like geodesic in  Lorenzian space time  (which corresponds to a tachyonic mode), the spatial and time like directions orthogonal to the tangent to the geodesic may conspire to give such degeneration.  

The problems pointed out in the previous paragraph however, do not take place for time like geodesics in a Lorenzian space time. The solution $v^\alpha=(t-t_0)u^\alpha$ is in $\mathbb{J}[t_1]$. Therefore, the space $\mathbb{J}[t_1]^{\perp}$ never contains the time like direction. This means that
 $g|_{\mathbb{J}[t_1]^{\perp}}$ is becomes spatial, and
\be\lb{espacializar}
g|_{\mathbb{J}[t_1]^{\perp}}=h|_{\mathbb{J}[t_1]^{\perp}},
\ee 
with $h_{ab}$ the transverse metric defined in (\ref{espasial}). The focal index then may be rewritten as
\be\lb{ifocalo}
i_{focal}(\gamma)=\sum_{t\in [a,b]} \text{signature} (h|_{\mathbb{J}[t]^{\perp}}).
\ee
If it is assumed that the full metric is non degenerate, then this restriction will also be non degenerate and $n_-(\widetilde{g})=0$ due to the spatial property. Thus, for time like geodesics
\be\lb{equaliti} 
i_{focal}(\gamma)=i_{geom}(\gamma).
\ee 
Although the definition $i_{geom}(\gamma)$ seems more natural for studying geodesics, the index $i_{focal}(\gamma)$ is more adequate for studying geodesic perturbations. Thus, this identification will be helpful for understanding the behaviour of conjugate points under small perturbations of $\gamma$. These type of indices were extensively studied in \cite{piccione1}-\cite{piccione14}.
\\

\emph{The case for null geodesics}
\\

For null geodesics, the analysis of conjugate points is slightly different as for the time like case described above.
One difference is that, given a null geodesic with tangent vector $k^\alpha$ and a deviation vector $v^\alpha$, the naive orthogonal condition $v^\alpha k_\alpha=0$ may be satisfied by fixing $v^a= c k^a$ with $c$ a constant, as $k^a$ has zero norm. This orthogonality condition does not remove the unwanted solutions parallel to the geodesic, as $k^\alpha$ is parallel and perpendicular to itself. 

 Consider as before a congruence of  null geodesics $x^\alpha(\tau, s)$ with $\tau$ the evolution parameter. The notation  $\tau$ is employed here in order to distinguish from the parameter $t$ employed in the time like case. Given an initial point $p$ corresponding to the initial parameter $\tau_0$, an initial orthonormal basis $e_+$, $e_-$ and $e_a$ with $a=1,2,..,n-2$ may be constructed for which
the metric at $p$ becomes
 \be\lb{orto4}
g=-2 e_+\otimes e_-+\delta_{ab} e^a \otimes e^b.
\ee
These coordinates are inspired in the null coordinates for special relativity. Clearly both $e_\pm$ are null vectors. The tangent vector $k^\alpha$ to $\gamma$ satisfies the geodesic equation $\nabla_{k} k=0$, thus it is parallel translated along the curve. The choice
$e_+(\tau_0, s)=k^\alpha(\tau_0, s)\partial_{\alpha}$ may be employed, with  the other components $e_-(\tau_0, s)$ and $e^a(\tau_0, s)$ only constrained by (\ref{orto4}). If this initial basis is parallel transported along the geodesic $\gamma_s$, that is, if $k^\alpha \nabla_\alpha e_a=k^\alpha \nabla_\alpha e_-=k^\alpha \nabla_\alpha e_+=0$ then 
\be\lb{orto3}
g=2 e_+\otimes e_-+\delta_{ab} e^a \otimes e^b,\qquad e_+(\tau, s)=k^\alpha(\tau, s)\partial_\alpha,
\ee
for all values of $\tau$ in a neighburhood of $\tau_0$.
In this case $e_+$ represents the tangent (and orthogonal) vector to the null geodesic at every value of $\tau$. The $n-2$ spatial dimensional metric \be\lb{catch}h_{ab}=\delta_{ab} e^a \otimes e^b,\ee is transverse to both the directions spanned by $e_+$
and $e_-$. The deviation vector $v^a$ satisfies the Lie condition ${\cal L}_{e_+} v=[v,e_+]=0$, which is equivalent to
$$
\nabla_{e_+} v = \nabla_v e_+.
$$
By expanding $v$ into the parallel transported basis as $v=v^+e_++v^- e_-+v^a e_a$, and by defining the indices $A=+,-, a$ the last formula becomes
\be\lb{west}
k^\alpha e_A \partial_\alpha v^A=v^\alpha \nabla_\alpha e_+.\ee
On the other hand, from the fact that $g(e_+, e_+)=0$ it is seen that
$g(e_+, \nabla_\alpha e_+)=0$. By expanding $ \nabla_\alpha e_+$ as
$$
\nabla_\alpha e_+=(\nabla_\alpha e_+)^+ e_++(\nabla_\alpha e_+)^- e_-+(\nabla_\alpha e_+)^a e_a,
$$
and by inserting this into the identity $g(e_+, \nabla_\alpha e_+)=0$, it is found that
$$
(\nabla_\alpha e_+)^-=0.
$$
In these terms, it is calculated from (\ref{west}) that
\be\lb{prelo}
\partial_\tau v^-=0.
\ee
Thus $v^-$ is constant along the geodesic and if initial conditions of the form (\ref{init}) are employed, then $v^{-}=0$ during all the evolution. Given the deviation vector $v$ in arbitrary coordinates, with components $v^\alpha$, its transversal  components $v_t$ can be found by use of the projection
\be\lb{transversal}
v_t^\alpha=h^\alpha_\beta v^\beta,
\ee
with $h_{ab}$ the spatial $n-2$ dimensional metric defined in (\ref{catch}).
These components satisfy the transversal Jacobi equation
$$
\frac{d^2 v_t^a}{d\tau^2}=-R^a_{+c+}  v_t^c,
$$
where the sign $+$ means the projection along $k^\mu$. This derivation is given in \cite{blau}.

Based on the analysis given above, the space of solutions of the Jacobi equation
 \be\lb{jacos}
\frac{d^2 v^a}{d\tau^2}=-R^a_{+c+}  v^c,
\ee
becomes of interest. However, the subtleties described for the time like case also arise here. In finding this equation, at the step (\ref{prelo}) above, the condition $v^-=0$ was employed. However the formula (\ref{jacos}) with the initial conditions (\ref{inito}) alone does not imply that $v^-=0$ for a generic solution. Nevertheless, the solutions with a component $v^-$ do not give rise a conjugate point. Neither does the solution $v^\alpha=(\tau-\tau_0)k^\alpha$. 
Thus, the multiplicity of a conjugate point for a null geodesic is at most $n-2$ and the space of non trivial Jacobi fields $\mathbb{J}[\tau]$ contains the directions $e_+$ and $e_-$. The natural spatial metric $h_{ab}$ for null geodesics is (\ref{catch}), which has $n-2$ dimensions. There is a further subtlety however, and it is that both $e_+$ and $e_-$ are null, thus they are parallel and perpendicular to themselves simultaneously. This may raise a question about wether they belong to $\mathbb{J}[\tau]$ or $\mathbb{J}[\tau]^\perp$ or to both. However as the directions  $e_+$ and $e_-$ do not give rise to conjugate points, they surely belong to $\mathbb{J}[\tau]$. At one hand, $g(e_+, e_+)=g(e_-, e_-)=0$. On the other hand $g(e_-, e_+)=1$.
Thus, these directions can not be included additionally in $\mathbb{J}[\tau]^\perp$ as they can be perpendicular to all the remaining spatial directions $e_a$ and to themselves, but the condition $g(e_-, e_+)=1$ states that these two directions are not orthogonal.
 
The focal and geometrical indices $i_{focal}$ and $i_{geom}$  may be defined by use of (\ref{igeom}) and (\ref{ifocal}) in the present case.  The focal index is
\be\lb{ifocala}
i_{focal}(\gamma)=\sum_{t\in [a,b]} \text{signature} (g|_{\mathbb{J}[t]^{\perp}}),
\ee
and the restriction is clearly spatial, as the null components $e_+$ and $e_-$ are not in $\mathbb{J}[t]^{\perp}$. The identification (\ref{equaliti}) also remains valid for null geodesics. The only difference is that the maximum multiplicity for a conjugate point is $n-2$ in the null case, while it is $n-1$ in the time like case.

It should be emphasized that the Jacobi equation may be interpreted in terms of a second variation of a length functional. This is an involved task, details can be found in \cite{hawking}-\cite{erlich} but this line will be not pursued further here.

\subsection{A conserved two form for the Jacobi problem}

The Jacobi problem admits a conserved two form $\omega$, which will be described and employed below.  This form already appear in the works \cite{robbin2}, \cite{helfer}-\cite{helfer2}.
These references are related, in particular,  to Morse theory. This theory was in fact applied to riemannian geodesics in the seminal mathematical works \cite{Milnor}-\cite{Bott}
and further employed for geodesics in semi-riemannian geometry in \cite{Beem}-\cite{Uhlenbeck}, a subject which is complicated due to non positivity of the space time metric.

 In order to describe the conserved form $\omega$, note that the Jacobi equations for time like and null geodesics were reduced in the previous subsection to the form
\be\lb{jaco}
\frac{d^2 v^ a}{du^2}=r^a_{\;b}(u) v^b,
\ee
with $r^a_{\;b}$ a term constructed in terms of the curvature $R^a_{bcd}$
and in terms of the tangent vector $u^a$ or $k^a$ of the geodesic under study. In addition, the parameter $u$ employed in (\ref{jaco}) is an unified notation denoting either the $t$ parameter for time like geodesics
or the parameter $\tau$ for null geodesics employed above. Since the curvature $R_{abcd}=\eta_{ae}R^e_{bcd}$ constructed in terms of the Levi-Civita connection of $g_{\mu\nu}$ satisfies the identity
$R_{abcd}=R_{cdab}$, it follows that the curvature term $r_{ab}=\eta_{ac}r^c_b$
is symmetric with respect to the interchange of its indices.  As discussed in the previous subsection, for studying conjugate points to a given point $p$ in $M$, it is customary to impose the initial conditions
\be\lb{init}
 v^\alpha(u_0)=0,\qquad \dot{v}^a(u_0)=I^a,
\ee
where the initial velocities $I^a$ are real arbitrary parameters. Given the differential problem (\ref{jaco}) the conserved form is given by
\be\lb{simple}
\omega((v,\dot{v}), (w,\dot{w}))=g(v,\dot{w})-g(w,\dot{v}).
\ee
where $g=\eta_{ab} e^a\otimes e^b$ with $e_a$ being the chosen parallel propagated orthogonal basis along $\gamma$, $e^a$ the corresponding dual basis,  and $\eta_{ab}$ the standard Minkowski metric in $n$ dimensions. Both $v$ and $w$ 
are vector fields in $T_rM$, the dot denotes the derivative with respect to $u$. This definition shows that $$\omega: P_r\times P_r\to R, $$
where the following phase space
\be\lb{fase}P_r=T_rM_r\oplus T_rM=\{(v, \dot{v}) | v \in T_r M \;\; \text{and} \;\;\dot{v}\in T_r M\},\ee
has been introduced. Here $r$ is any point of the geodesic $\gamma$ under study.

The expression (\ref{simple}) for $\omega$ is not the most common in the literature and so, it may be useful to find another equivalent definitions. 
Note that  the phase space  $P_r$ defined in (\ref{fase}) is composed by the direct sum of two copies of $T_rM$.
For this reason, it is convenient to introduce a copy $f_a$ of the $n$-bein basis  $e_a$  of $T_rM$ described above. Despite being a copy, the new vectors $f_a$ live in a complementary space to the one spanned by $e_a$ in the sense that $$e_a(e^b)=f_a(f^b)=\delta_{a}^b,$$ but
$$e_a(f^b)=f_a(e^b)=0.$$ In the formalism of phase space, instead of expanding a deviation vector $v$ of $\gamma$ and its velocity $\dot{v}$ as
$$
v=v^a e_a,\qquad \dot{v}=\dot{v}^a e_a,
$$
one assigns to them a $2n$-dimensional vector $V\in P$ given by
\be\lb{newvect}
V=v^a e_a+\dot{v}^a f_a.
\ee
In these terms, the form (\ref{simple}) becomes
\be\lb{simple2}
\omega=\eta_{ab} e^a\wedge f^b,
\ee
and in fact, given two vectors $V$ and $W$ in $P_p$ expanded as in (\ref{newvect}), it is concluded directly from (\ref{simple2}) that 
$$
\omega(V, W)=\eta_{ab} v^a\dot{w}^b-\eta_{ab}\dot{v}^a w^b.
$$
This proves that (\ref{simple2}) and (\ref{simple}) are two expressions of the same object. 

The expression of the form $\omega$ may take even a more familiar form by making the redefinition $e^0\to -e^0$ while leaving all the spatial components of $e^a$ and all the components $f^a$ unchanged. In this new basis, the most familiar expression
\be\lb{simple3}
\omega=\delta_{ab} e^a\wedge f^b,
\ee
it is obtained, which is typical in the literature.

For any two solutions $v$ and $w$ of the Jacobi equation 
(\ref{jaco}) it can  be deduced that 
\be\lb{closo}
\frac{d\omega}{du}((v,\dot{v}), (w,\dot{w}))=0.
\ee
This follows directly from  (\ref{jaco}) and (\ref{simple}) and from the fact that $r_{ab}=r_{ba}$. If in addition, the initial conditions (\ref{init}) are employed, then it follows that
\be\lb{simplel}
\omega((v,\dot{v}), (w,\dot{w}))=0,
\ee
for every value of $u$. This is valid on any causal geodesic $x^\mu(u, s)$.

The formula (\ref{simplel}) has some important consequences.
First, by use of the definition  (\ref{simple}) it may be written as
\be\lb{wrass}
g(v, \dot{w})=g(w, \dot{v}),
\ee
with $v$ and $w$ two solutions of (\ref{jaco}) corresponding to the conditions (\ref{init}). 
If, at a parameter $u_1$, a conjugate point is of multiplicity $k$ appears, there are $k$ linearly independent Jacobi vectors $v_i$
which vanish at $u_1$. Their corresponding velocities $\dot{v}_i$ should be linearly independent. If this were not the case, then there is a combination $\dot{c}=\alpha^i\dot{v}_i$ with $i=1,..,k$
which vanish at $u_1$. The linear combination $c=\alpha^i v_i$ also vanish at $u_1$. As this is a solution of the Jacobi equation, which is linear and of second order, it is clear that $c=0$. But this contradicts the linear independence of $v_i$. Thus the velocities $\dot{v}_i$ are linearly independent. The last formula implies that
\be\lb{ortogonaliza}
g(w(u_1), \dot{v}_i(u_1))=0,\qquad i=1,..,k,
\ee
with $w(u_1)$ any vector belonging to $\mathbb{J}[u_1]$, that is, the space of non trivial Jacobi fields at $u_1$. 

It is not difficult to convince oneself that (\ref{ortogonaliza}) implies that the velocities
$\dot{v}_i(u_1)$ belong to $\mathbb{J}[u_1]^\perp$. But this should be analysed with care, due to the fact that $g$ is not riemannian. For time like geodesics, the solution $w^\alpha(t_1)=(t_1-t_0)u^\alpha$ described in the previous section may be employed to show that
$\dot{v}_i(u_1)$ is spatial due to (\ref{ortogonaliza}). For spatial $w(u_1)$, the metric $g_{ab}$ in (\ref{ortogonaliza}) becomes riemannian, and it is clear that $\dot{v}_i(u_1)$ belong to $\mathbb{J}[u_1]^\perp$.
For null geodesics, a vector $w(u_1)$ proportional to $e_+$ may be employed in order to show that $\dot{v}_i(u_1)$ does not contain a component proportional to  $e_-$, since
$g(e_-, e_+)=1$. The same type of argument can be used to show that $\dot{v}_i(u_1)$  does not have a component proportional to $e_+$. For spatial $w(u_1)$, the metric $g_{ab}$ in (\ref{ortogonaliza}) becomes riemannian. This means that the velocities are in $\mathbb{J}[u_1]^\perp$ in the null case. Together with the linear independence, these facts imply the following lemma \cite{piccione1}.
\\

\emph{Lemma 1:} Given a conjugate point $q$ to $p$ with multiplicity $k$ corresponding to the evolution parameter $u_1$ on a causal geodesic $\gamma$, there are $k$ linearly independent Jacobi fields $v_l$ with $l=1,2,..,k$ which vanish at $u_1$. Their velocities $\dot{v}_l(u_1)$ at this point constitute a basis for
$\mathbb{J}[u_1]^\perp$.
\\

The presence of the conserved form $\omega$ have several other consequences. The phase space $P$ is $2n$ dimensional. A $n$-dimensional subspace $L$ of $P$ is called \emph{lagrangian} if the restriction of $\omega|_{L}$ to this space is zero, namely
\be\lb{ls}
 \omega(e_i, e_j)=0,\qquad \forall\; e_i, e_j\in L.
\ee
For the Jacobi problem, the subspaces $L_1=\text{span}\{e_i\}$ corresponding to the Jacobi vectors or $L_2=\text{span}\{f_i\}$ corresponding to their velocities are of course lagrangian.
In addition, two lagrangian subspaces $L_1$ and $L_2$ are said to be complementary if $L_1\oplus L_2=P$. For any pair of such complementary lagrangians, there exist a basis $e_i\in L_1$ and $f_i\in L_2$ such that
 $$
 \omega(e_i, f_j)=\delta_{ij}.
 $$
And any other complementary space $L_c$ to $L_1$ if of the form $L_c=\text{span}\{g_i\}$ with
\be\lb{mitre}
g_i=f_i+C_{ij} e_j.
\ee
The matrix $C_{ij}$ has to be constrained by the lagrangian condition $\omega(g_i, g_j)=0$. This condition, together with $\omega(e_i, e_j)=\omega(f_i, f_j)=0$
and $\omega(e_i, f_j)=\delta_{ij}$ implies that $C_{ij}=C_{ji}$. In other words, the space of lagrangians complementary to any specific $L_1$ is described by the space of symmetric $C_{ij}$ matrices.

The space of all the lagrangian subspaces of $P$ will be denoted as $G(n, P)$ and is called Grassmanian lagrangian. Several of its properties are studied in the literature, see for instance \cite{piccionelect}. Some very basic properties are collected in the appendix at the end of the text. 

\subsection{Lagrangian planes determined by the solutions of the Jacobi equation}
Equipped with phase space formalism, consider again the Jacobi equation \eqref{jaco}. Instead of studying particular initial conditions (\ref{init}),  the attention will be focused now on the space of all possible initial conditions of this type. Any solution $v^a$ of (\ref{jaco}) can be parametrized in terms of these conditions (\ref{init}) as follows
\be\lb{gensol1}
v=A^a_b(u) \dot{v}^b(u_0) e_a(u), 
\ee
with $e_a$ the n-bein basis parallel transported along $\gamma$ introduced above, $\dot{v}^b(u_0)$ arbitrary initial velocities, and with the matrix $A^a_b(u)$ defined through
the second order differential equation
 \be\lb{jaco3}
\frac{d^2 A^a_b}{du^2}=-r^a_{c}  A^c_b,\qquad A^a_b(u_0)=0,\qquad \frac{dA^a_b(u_0)}{du}=\delta_b^a.
\ee
This characterization of $A^a_b(u)$ follows directly from the Jacobi equation and the aforementioned initial conditions.
Once the matrix $A^a_b(u)$ has been found, the space of all solutions corresponding to the initial conditions (\ref{init}) is obtained as
\be\lb{gensol1}
V=A^a_b(u) \dot{v}^b(u_0)e_a(u)+\dot{A}^a_b(u) \dot{v}^b(u_0)f_a(u).
\ee
The formula (\ref{simplel}) shows that $\omega(V, W)=0$ for every  pair of such vectors $V$ and $W$, if they correspond to the initial conditions (\ref{init}). Thus, the definition (\ref{ls}) shows that the space of all the vectors of the form (\ref{gensol1}) constitute a lagrangian subspace $L_1(u)$ of $P$ for every value of $u$. From the initial conditions (\ref{init}) it is seen that
$L_1(u_0)= 0\oplus T_p M$. 

Given  $L_1(u)$ as above, a  complementary lagrangian space $L_2(u)$ may be constructed for every value of $u$, by searching the solutions of the Jacobi equations (\ref{jaco}) with the initial conditions
\be\lb{init2}
 v^a(u_0)\in \mathbb{R}^n,\qquad \dot{v}^a(u_0)=0.
\ee
Clearly, this initial set is complementary to the one defined in (\ref{init}), since it corresponds to non zero initial vectors with zero velocities. The  corresponding solutions are characterized by the following formula
$$
v=B^a_b(u) v^b(u_0)e_a(u),
$$
with a matrix $B^a_b(u)$ satisfying the differential system
 \be\lb{jaco4}
\frac{d^2 B^a_b}{du^2}=-r^a_{c}  K^c_b,\qquad B^a_b(u_0)=\delta^a_b,\qquad \frac{dB^a_b(u_0)}{du}=0.
\ee
All the solutions of the complementary Jacobi problem
\be\lb{gensol2}
v(u)=B^a_b(u) v^b(u_0)f_a(u),
\ee
are obtained by varying the initial displacements $v^b(u_0)$ over $\mathbb{R}^n$. Corresponding to these solutions, a phase space vector
\be\lb{genso22}
V_c=B^a_b(u) v^b(u_0)e_a(u)+\dot{B}^a_b(u) v^b(u_0)f_a(u),
\ee
is assigned. As before, the space of all these solutions is a lagrangian subspace of $P$, denoted as $L_2(u)$. The conditions (\ref{init2}) imply that
$L_2(u_0)= T_p M\oplus 0$. 

 Clearly, the initial lagrangian subspaces $L_1(u_0)=0\oplus T_p M$ and $L_2(u_0)= T_p M\oplus 0$ are complementary, since  obviously $L_1(u_0)\oplus L_2(u_0)=P_p$.
 
 Note that the vector field $V$ in (\ref{gensol1}) and the complementary vector $V_c$ in (\ref{genso22}) can be expressed as
$$
V=\dot{v}^a(u_0) F_a(u),\qquad V_c=v^a(u_0) E_a(u),
$$
where the following $u$ dependent $n$-bein basis
\be\lb{baso}
F_a(u)=A^b_a(u)e_b(u)+\dot{A}^b_a(u) f_b(u),\qquad E_a(u)=B^b_a(u)e_b(u)+\dot{B}^b_a(u) f_b(u),
\ee
has been introduced. In this basis the simple description \be\lb{spam}L_1(u)=\text{span}\{F_a(u)\},\qquad
L_2(u)=\text{span}\{E_a(u)\},\ee is achieved. It can be easily seen that $E_a(u_0)=e_a(u_0)$
and $F_a(u_0)=f_a(u_0)$. At  $u_0$ by construction $\omega(e_a(u_0), f_b(u_0))=\delta_{ab}$
and since this value is conserved during the $u$ evolution by (\ref{closo}) it follows that $\omega(E_a, F_b)=\delta_{ab}$
for every value of $u$. Thus, $E_a(u)$ and $F_a(u)$ are complementary basis for any choice of $u$ and therefore the lagrangians $L_1(u)$
and $L_2(u)$ are complementary for every value of $u$.

Having introduced  the basis (\ref{baso}) for $P$, the next step is to obtain some formulas that will be useful for reproducing the geometrical index $i_{geom}(\gamma)$ and the focal index $i_{focal}(\gamma)$ defined in (\ref{igeom})-(\ref{ifocal}) in the phase space formalism. For achieving this, consider a fixed arbitrary value of the evolution $u_f$ not corresponding necessarily to a conjugate point. The value of  $F_a(u)$ defined in (\ref{baso})  can be related on general grounds to the ones at $u_f$ by some combination of the form
\be\lb{lijas}
F_a(u)=F_a(u_f)+L_{ab}(u_f, u) E_b(u_f).
\ee
By inserting the last expression into  $\omega(E_a(u), E_b(u_f))$  and by taking into account the complementary property of $E_a(u)$ and $F_b(u)$ namely, that
$\omega(E_a, F_b)=\delta_{ab}$, the unknown quantity $L_{ab}$ may be calculated explicitly as
\be\lb{lij}
L_{ab}(u_f, u)=\omega(F_a(u), F_b(u_f)).
\ee
The point of introducing the quantity $L_{ab}$ is that the intersection of $L_1(u)$ with the initial lagrangian plane $L_1(u_0)$ defined by the initial conditions of the form (\ref{init}) can be characterized by the zero eigenvalues of the matrix $L_{ab}(u_0, u)$. This can be explained as follows. The initial lagrangian  is $L_0=\text{span}\{f_a(u_0)\}$, while the evolved one is  
$$
L_1(u)=\text{span}\{F_a(u)\}=\text{span}\{f_a(u_0)+L_{ab}(u_0, u) e_{b}(u_0)\}.
$$
This means that a  vector $V\in L_1(u)$ given by
\be\lb{nulo}
V=\dot{v}^a(u_0) F_a(u),
\ee
with $\dot{v}^a(u_0)$ the initial velocities,  will also belong to $L_1(u_0)$ and therefore to the intersection $L_1(u)\cap L_1(u_0)$ if and only if
\be\lb{nulo2}
\dot{v}^a(u_0) L_{ab}(u_0, u)=0.
\ee
The last condition implies that $\dot{v}^a(u_0)$ must be an eigenvector of $L_{ab}(u_0, u)$ corresponding to the eigenvalue $\lambda=0$. Therefore
$$
\text{dim}(L_1(u)\cap L_1(u_0))=\text{multiplicity of the eigenvalue} \;\lambda=0\; \text{of}\; L_{ab}(u_0, u).
$$
This last conclusion may be paraphrased as follows. For a given value of $u$, the matrix $L_{ab}(u)$ has $n$ eigenvalues $\lambda_i(u)$, possibly repeated, and not necessarily vanishing. If there is a value $u_1$ such that
one these values is zero with multiplicity $k$, then this corresponds to a conjugate point $q$ to $p$ on $\gamma$ with the same multiplicity. 

Another important quantity is the derivative of $L_{ij}$ with respect to $u$, which will be denoted here by $Q_{ij}$ in order to facilitate comparison with the literature. This derivative is found choosing $u=u_f+\delta u$ with $\delta u$ infinitesimal, and by realizing that the definition (\ref{lij}) implies that $L_{ij}(u_f, u_f)=0$. The result is
\be\lb{derivv}
Q_{ij}=\frac{dL_{ij}(u_f, u)}{du}=\omega(\dot{F}_i(u_f), F_j(u_f)).
\ee
On the other hand, the first equation in (\ref{baso})
shows that
\be\lb{paso}
\dot{F}_i=u^\alpha \nabla_\alpha F_i=\dot{A}^k_i(u)e_k+\ddot{A}^l_i(u) f_l=\dot{A}^k_i(u)e_k+r^l_c A^c_i f_l,
\ee
as the basis $e^a$ and $f^a$ are parallel translated along $\gamma$. Note that in the last equality the equation (\ref{jaco3}) has been employed. By inserting the last expression and the first (\ref{baso}) into (\ref{derivv}) and by taking
the definition of $\omega$ given in (\ref{simple2}) into account, it is concluded that
\be\lb{derivv2}
Q_{ij}(u)=\eta_{ab} \dot{A}^a_i \dot{A}^b_j-r_{ab}A^a_i A^b_j.
\ee
It may be more illuminating to write the last formula as
\be\lb{paradisio}
Q_{ij}(u)=[A_i^a(u)\;\; \dot{A}^a_i(u)]
\left[
\begin{array}{cc}
-r_{ab}  & 0    \\
0  &   \eta_{ab}   
\end{array}
\right]
\left[
\begin{array}{c}
 A_i^b(u)\\
 \dot{A}_i^b(u) 
\end{array}
\right].
\ee
Since $v^a(u)=A^a_b(u) \dot{v}^b(u_0)$ and $\dot{v}^a(u)=\dot{A}^a_b(u) \dot{v}^b(u_0)$, it is seen that the quantities $Q_{ij}$ define a symmetric tensor $Q$
acting on $P\times P$ given by
\be\lb{simten}
Q=r\oplus g,\qquad r=r_{ab}e^a\otimes e^b,\qquad g=\eta_{ab}f^a\otimes f^b.
\ee
This is the desired expression for the derivative $Q_{ij}$ in an arbitrary basis \cite{helfer}.

\subsection{Definition of the Maslov index}

After this characterization of $L_{ij}$ and its derivative $Q_{ij}$,  a new expression for the focal and geometrical indices $i_{focal}(\gamma)$ and $i_{geom}(\gamma)$ may be found. 
The importance of this expression, to be derived below, is that its properties under geodesic deformations can be understood in terms of classical mathematical literature.
For obtaining it, consider again the Jacobi problem (\ref{jaco}), with initial conditions (\ref{init}) at a point $p$ corresponding to the parameter $u_0$.
Assume that there is a point $q$ corresponding to a parameter $u_1$ such that there are $k$ non trivial initial velocities $\dot{v}_l(u_0)$ with $l=1,..,k$ for which the zero eigenvalue condition (\ref{nulo2}) is satisfied. These formulas show that, for fixed $l$, the quantities $\dot{v}^i_l(u_0)$ with $i=1,..,n$ constitute a zero eigenvector of $L_{ij}(u_0, u)$. Associated to this eigenvector, a phase space vector
 (\ref{nulo}) may be constructed, which can be explicitly worked out from (\ref{baso}) as
\be\lb{nulo}
V_l=\dot{v}^a_l(u_0) J^b_a(u)e_b+\dot{v}^a_l(u_0) \dot{J}^b_a(u) f_b.
\ee
This corresponds to a deviation vector $v_l^a(u)=\dot{v}_l^b(u_0) J^a_b(u)$ with velocity $\dot{v}_l^a(u)=\dot{v}_l^b(u_0) \dot{J}^a_b(u)$. At the  conjugate point $q$ 
$$
v_l^a(u_1)=\dot{v}_l^b(u_0) J^a_b(u_1)=0,
$$
and therefore $V_l(u_1) \in L_1(u_1)\cap L_1(u_0)$, as discussed below (\ref{nulo}). In this situation, the space of non trivial Jacobi fields $J[u_1]\in T_qM$ becomes $n-k$ dimensional, since there are $k$ Jacobi fields
are vanishing at $q$. The orthogonal complement $J[u_1]^\perp\in T_q M$ is then $k$ dimensional. On the other hand, the vectors (\ref{nulo}) at $u_1$ become
\be\lb{nulon}
V_l(u_l)\simeq \dot{v}_l(u_1)=\dot{v}^a_l(u_0) \dot{J}^b_a(u) f_b.
\ee
Here $\simeq$ denotes that these quantities are isomorphic, as $f_a$ are copies of $e_a$. The lemma proved below formula (\ref{ortogonaliza})
shows that the vectors (\ref{nulon}) conform a basis of $J[u_1]^\perp$. The restriction of $Q=\dot{L}$ on the space spanned by the vectors  (\ref{nulon}) can be obtained from (\ref{simten}), the result is simply
$$
Q|_{L_1(u)\cap L_1(u_0)}=g|_{J[u_1]^\perp},
$$
 with $g$ the full space time metric. By comparing this with the definitions (\ref{igeom}) and (\ref{ifocal}) of  $i_{geom}$ and $i_{focal}$
 it follows, for causal geodesic curves in  the space time ($M$, $g$), that
 \be\lb{indice}
 i_{focal}(\gamma)=i_{geom}(\gamma)=\sum_{u\in [a,b]} \text{signature}
(Q|_{L_1(u)\cap L_1(u_0)}).
 \ee
The quantity
\be\lb{Maslov}
\mu(\gamma, L_1(u_0))=\sum_{u\in [a,b]} \text{signature}
(Q|_{L_1(u)\cap L_1(u_0)}),
\ee
is known as the Maslov index for $\gamma$ at the interval $[a,b]$ \cite{maslov}, \cite{robbin1}. Here it is assumed that $a<u_0<b$ and that $a$ and $b$ do not represent a conjugate point to the initial point $p$
corresponding to $u_0$. The possibility of the borders containing conjugate points will be described in the next section.
Then (\ref{indice}) is equivalent to
\be\lb{identico}
i_{focal}(\gamma)=i_{geom}(\gamma)=\mu(\gamma, L_1(u_0)).
\ee
This formula identify the quantities of interest for geodesics in terms of the Maslov index. This identification will be employed in the next sections in order to prove some homotopy properties about the focal and geometrical indices for causal geodesics in lorenzian geometry.

\section{The behavior of conjugate points by moving geodesics}

\subsection{The  properties of the Maslov index under homotopy}

At first sight, the identification (\ref{identico}) may look a bit convoluted. The first two quantities are easy to define, but the last one is more involved. The advantage of this identification is that the 
homotopy properties of the Maslov index are well studied \cite{robbin1}. This may be helpful for studying the behaviour of the conjugate points when moving alone
the congruence  of geodesics $\gamma_s$ containing a given one $\gamma$. Before going about this topic, it is mandatory to show the link between 
the formula (\ref{simten}) with the ones described in the classical literature of the subject. As this identification is lengthy, this is done in the appendix.
In particular, the formulas for $Q_{ij}$  and the Maslov index expressions obtained in \cite{salamon} are reproduced exactly.

Several conclusions may be drawn by use of that reference. These authors define the Maslov index of the lagrangian path $L_1(u)$ with respect to any chosen lagrangian plane $V$
as
$$
\mu(L_1(u), V)=\frac{1}{2}\text{signature}
(Q(a)|_{L_1(u)\cap V})+\frac{1}{2}\text{signature}
(Q(b)|_{L_1(u)\cap V})+\sum_{u\in (a,b)} \text{signature}
(Q(u)|_{L_1(u)\cap V}).
$$
This definition is not restricted to a problem of geodesics, and in fact the Maslov index has several applications in other context such as Quantum Mechanics.
The only subtle difference with our definition is the contribution of the endpoints $a$ and $b$. But this can be understood, as
 the authors of \cite{salamon} deals with a more general context, not limited to 
lagrangian paths $L_1(u)$ arising by a problem of geodesics. In addition, the endpoint $b$ will be chosen such that it is not a conjugate point
to $a$, thus it will give no contribution to the index. irst, the values of $u$ where $L_1(u)\cap V$ is non trivial are called crossings. A crossing is called regular if the quadratic form $Q|_{L_1(u)\cap V}$ is non singular.
 The first two homotopy properties are the following.
\\

\emph{Homotopy property:} Given two homotopic lagrangian curves $L_1(u)$ and $\widetilde{L}_1(u)$ with the same endpoints and with regular crossings only have the same Maslov index. Furthermore any lagrangian path $L_1(u)$ is homotopic to one with only regular crossings.
\\

Another important property is related to path catenation.
\\

\emph{Catenation property:} Given a generic Grassmanian $G(n)$ and a curve $\gamma:[a,b]\to G(n)$, then the following additivity property is valid
$$
\mu(\gamma, \gamma(a))=\mu(\gamma|_{[a,c]},\gamma(a))
+\mu(\gamma_{[c,b]},\gamma(c)),
$$
with $c\in (a,b)$.
\\

The last fundamental property for our purposes is the following one.
\\

\emph{Homotopy property 2:} Given two lagrangian curves $L_1(u)$ and $\widetilde{L}_1(u)$ with the same endpoints are homotopic if and only if they have the same Maslov index.
\\The properties described above define the Maslov index in full generality for lagrangian paths, by assigning to a path with singular crossings with $L_1(u_0)$
the Maslov index of an homotopic deformed path with regular crossings, which always exists due to Proposition 1.
\\

These homotopy properties will be fundamental for the proof of the proposition given in the following section.

\section{The continuity property of conjugate points along $\gamma_s$}

All the statements given so far are related to the conjugate points of a given geodesic $\gamma$.
However, nothing has been said about the behavior of these conjugate points when moving along the geodesics $\gamma_s$ of a congruence  in which $\gamma$ is embedded in.
The homotopy property presented in the previous subsection may be useful for doing so, and we turn the attention over this subject below. 

Consider a geodesic $\gamma$ embedded in a congruence $\gamma_s$ in such a way that $\gamma_0=\gamma$. Take a point $p$ corresponding
to the parameter value $u_0$ and assume that there is conjugate point  $q$ to $p$ along $\gamma$ at a parameter value $u_1$. Without loss of generality
$q$ may be assumed to be the first conjugate point to $p$. Then, for causal curves, it may be shown that there are no conjugate points to $p$ in an interval
$[u_0, u_0+\epsilon]$ or $[u_1, u_1+\epsilon]$, for $\epsilon$ small enough. In other words, the set of conjugate points $q_i$ to $p$ is discrete. This is the statement
of the following lemma \cite{oneil}. 
\\

\textit{Lemma 2:} Given a point $p$ in a causal geodesic corresponding to a parameter $u_0$ there is an $\epsilon$ such that there is no conjugate
points to $p$ for $u<u_0+\epsilon$. If a conjugate points $q$ appears at a parameter $u_1$ then there is an $\epsilon$ such that there
is no other conjugate point for $u<u_1+\epsilon$. 
\\

\begin{proof} This can be seen as follows.
At a conjugate point $q$ of multiplicity $k$ there are $k$ vanishing Jacobi fields $v_l$ with $l=1,..,k$ and $n-k$ non vanishing vectors one $v_l$ with $l=k+1, .., n$.
Construct the new vector fields
$$
\eta_l(u)=v_l,\qquad l=k+1,.., n
$$
$$
\eta_l(u_1)=\dot{v}_l(u_1),\qquad \eta_l(u)=\frac{v_l(u)}{u-u_0}, \qquad l=1,..,k.
$$
At the parameter $u_1$ the first vectors are a basis of $\mathbb{J}[u_1]$ and the second are a basis of $\mathbb{J}[u_1]^\perp$. As the metric $g$ is non degenerate on $\mathbb{J}[u_1]$ for causal geodesics, it follows that $\mathbb{J}[u_1]\oplus \mathbb{J}[u_1]^\perp=\mathbb{R}^n$. These vectors are continuous around $u_1$ and constitute a basis of $\mathbb{R}^n$ for $u$ close to $u_1$. Thus, for $u$ close to $u_1$ there is no conjugated point, as this would require $\eta_l$ to be a basis of $R^m$ with $m<n$ which by continuity of the determinant $\det((\eta_l)_{l=1}^n)$ is impossible. 
\end{proof}

The lemma given above, combined with the definition of the Maslov index given in the previous sections, is crucial for the present purpose, which is the proof of the proposition given below.
\\

\emph{Proposition:} Given a space time $(M, g_{\mu\nu})$ consider a causal geodesic $\gamma$ with tangent vector $u^\alpha$ with two conjugate points $p$ and $q$, with $q$ in the causal future of $p$ (usually denoted by $q\in J_+(p)$). If this geodesic is embedded in a causal congruence $\gamma_s$  with tangent vector $u^\alpha_s$ where $s$ is a multidimensional parameter, in such a way that $u^\alpha_s\to u^\alpha$ and $\gamma_s\to \gamma$ when $s\to 0$.  Then, the following holds.
\\

a) There is an open neighbourhood $O$ around $p$ composed by points $p(s)$ belonging to $\gamma_s$ such that $p(s)\to p$ when $s\to 0$, in such a way that any of these points posses a conjugate point $q(s)$ in $J_+(p(s))$ along $\gamma_s$. 
\\

b) The map $h: O\to M$ defined by $h(p(s))=q(s)$, with $q(s)$ the first conjugate point to $p(s)$,
is continuous around $p$.
\\

\emph{Proof:} It is convenient to parametrize the causal geodesics $\gamma_s$ composing the congruence in local coordinates as $x^\mu(u, s)$ in such a way that $x^{\mu}(u, 0)$ corresponds to $\gamma$. For all the geodesics in the congruence, with $s$ small enough, find the general solution of the Jacobi equation the initial conditions (\ref{init})  at the parameter value $u_0$. There is no loss of generality with the particular choice of the initial parameter $u=u_0$. Then, all the possible Jacobi fields $v^\nu(u, s)$ are given by
$$
v(u, s)=A^a_b(u, s) \dot{v}^b(u_0, s) e_a(u, s),
$$
where the quantity $A^a_b(u, s)$ is defined in (\ref{jaco3}), and is clearly $s$ dependent since the curvature term $r_{ab}(s)$ is function of this parameter.  The initial velocities $\dot{v}^b(u_0, s)$ are varied over the possible real values.

The initial conditions $\dot{v}^b(u_0, s)$ may be chosen as continuous functions of $s$. The reason is that the last formula implies that if a conjugate point is reached for a given value of the initial condition $\dot{v}^b(u_0, s)$, then the same holds for the scaling $\dot{v}^b(u_0, s)\to \lambda \dot{v}^b(u_0, s)$, with $\lambda$ any real parameter, as the vanishing property will not be spoiled by this rescaling.  By considering the space of all  continuous initial conditions $\dot{v}^b(u_0, s)$ the space of all the possible initial conditions for the curves $x^\mu(u, s)$ at $u_0$ is covered.

Now, for a given geodesic $\gamma_s$, the Lemma 1 described above implies that, for some small neighbourhood  $[u_0-\epsilon(s), u_0+\epsilon(s)]$,
 there are no points  conjugate to the initial point $p(s)$ corresponding to the parameter $u_0$, that is, $p$ is the point located at $x^\mu(u_0, s)$. For $s$ taking values in a small compact interval $[-s_1, s_1]$, denote the smallest value of $\epsilon(s)$  as $\epsilon_s$. 
 Once this smallest value is defined, construct a continuous curve $\Gamma$ non necessarily geodesic, which starts at the point $p_0$ located at $x^{\mu}(u_0, 0)$ and ends at some point $r$ located at some coordinates $x^\mu(s_f, u_1)$ with $u_1$ in the interval $(u_0-\epsilon_s, u_0+\epsilon_s)$ and $s_f$ inside  the compact interval $[-s_1, s_1]$ defined above.  This curve is required to intersect all the curves $\gamma_s$  of the congruence for $s\leq s_1$, at least in the compact interval $[u_0-\epsilon_s, u_0+\epsilon_s]$. The intersection between $\Gamma$ with the curve $\gamma_s$ with coordinates $x^{\mu}(s, u)$ will happen at some point which is a function of $s$, and which will be denoted subsequently as $u_i(s)$. The curve $\Gamma$ is then parametrized by $s$, as it is composed by the points with coordinates $x^{\mu}(s, u_i(s))$. 
 
 In the same manner, construct another curve $\Delta$ joining a point $t$ with coordinates $x^\mu(s_f, u')$ in $\gamma_{s_f}$ with a point $q$ in $\gamma$ with coordinates $x^\mu(0, u'')$  with both parameters $u'>u_0+\epsilon_s$ and $u''>u_0+\epsilon_s$.
The curve $\Delta$ is assumed to intersect all the curves $\gamma_s$ in the congruence for $0\leq s\leq s_f$ with intersection points $v_i(s)$. In this manner, the points $p$ and $q$, both belonging to $\gamma$ become joined by the causal geodesic $\gamma$ or by the union curves $\Gamma_s\cup \gamma_s\cup \Delta_s$ with $s\leq s_1$. In particular, $\Gamma_{s_f} \cup \gamma_{s_f} \cup \Delta_{s_f}$  last ones connects $p$ with $r$, then $r$ with $t$ and finally $t$ with $q$.  Note that this composed curve is not necessarily geodesic, as the components $\Delta$ and $\Gamma$ are not restricted to be so. 

   The curves $\Gamma_s\cup \gamma_s\cup \Delta_s$ can be constructed in such a way that they are presented as  a one parameter deformation of  $\gamma$. To see this, assume without loosing generality that the intersection point function $u_i(s)$ is monotone and can be inverted locally as $s=s(u_i)$ along $\Gamma_s$ or $\Delta_s$. Then $\Gamma_s\cup \gamma_s\cup \Delta_s$ becomes a function of $u_0<u<u_1$. It depends on the final value of $s=s_f$, as this parameter define the endpoints of $\Gamma_s$ or $\Delta_s$. In addition $\Gamma_s\cup \gamma_s\cup \Delta_s\to\gamma$ when $s_f\to 0$, and both curves have the same endpoints $p$ and $r$.

  Once the curves $\Gamma$ and $\Delta$ have been constructed, the next task is to assign to every point of them the Jacobi vector field $v_\Gamma(u, s)=v(u, s)$ corresponding to the initial velocity $ \dot{v}^a(u_0, s)$ at the point of intersection $u_i(s)$ between $\Gamma$ and $\gamma_s$. The analogous procedure is made for the intersection points $v_i(s)$ in $\Delta$. On the other hand, at $\gamma_s$, the vector field is identified with the Jacobi field $v^a(u, s)$ at this curve.  By properties of linear differential equations, the vectors $v_\Gamma(u, s)$ and $v_\Delta(u, s)$ are continuous functions of the arguments since the initial conditions are of this type. These functions are Jacobi vectors, and formula (\ref{closo}) shows that the two form $\omega$ is conserved when evaluated on these vectors. Furthermore, the initial condition (\ref{init}) implies that its value is not only constant, but identically zero. Thus, along the curve  $\Gamma_s\cup \gamma_s\cup \Delta_s$ these fields constitute a continuous lagrangian path $L_\Gamma(u)\cup L_{\gamma_s}\cup L_\Delta(u)$ in $G(n, P)$.  The curve $L_{\Gamma}(u)\cup L_1(u, s_1)\cup L_{\Delta}(u)$ and $L_1(u, 0)$ in the grassmanian $G(n, P)$ have the same endpoints. For $s_1$ small enough, both curves are homotopic and the results presented in the previous section show that they have the same Maslov index.
  
     The equality between the Maslov indices between $L_\Gamma(u)\cup L_{\gamma_s}\cup L_\Delta(u)$ and $L_1(u, 0)$  shown above has an important consequence.  If there are no conjugate points between $p$ and $r$ in $\gamma$, the Maslov index is given only by the contribution of the initial point. On the other hand, the curves $\Gamma_s$ and $\Delta_s$ have been constructed intentionally for non containing conjugate points. Thus, the equality of Maslov indices between $\gamma$ and the curve $\Gamma_s\cup \gamma_s\cup \Delta_s$ implies that there is no conjugate point for $\gamma_s$ between 
the point $p(s)$ corresponding to the parameter $u_0$ and the intersection point between $\gamma_s$ and $\Delta_s$. If instead the point $r$ is allowed to move along $\gamma$ and happens to cross a conjugate point $q$ to $p$ with multiplicity $k$, then the equality of the Maslov index implies that a conjugate point $q(s)$ to $p(s)$ with the same multiplicity or $j$ conjugate points $q_i(s)$ of multiplicities $k_i$ with $i=1,.., j$ such that $k_1+..+k_j=k$ appear at the curve $\gamma_s$. The fact that $\Gamma_s\cup \gamma_s\cup \Delta_s\to\gamma$ when $s\to 0$ implies that  $q_i(s)\to q$ when $s\to 0$ in continuous fashion. This argument may be generalized without problems to a several parameter $\epsilon_i$ congruences, if they describe a ball with radius small enough. The proposition has then been proved (Q. E. D). 
\\

It should be mentioned that a result of this type was given in  \cite{gaowald}, but restricted to GR and to matter content satisfying the strong energy condition. A more general result was found in \cite{yo}. However, this work compare null geodesics with themselves, while the result presented here is not restricted  to this case alone, and  allows comparison between causal geodesics in general.
The application of the Maslov index generalize these statements to any matter content and any underlying gravitational model, for causal geodesics. This is an important feature. Its physical significance is that, given a null geodesic
which achieves two conjugate points, then time like geodesics which similar initial directions and with velocities close to light will also acquire a pair of conjugate points with the continuity property stated in the proposition.

\section{Discussion of the results}

The result presented in the previous section can be rephrased in less technical terms as follows. Given a causal geodesic $\gamma$ with two conjugate points $p$ and $q$, then there is an open neighbourhood around $p$ such that all the causal geodesics starting from points $p_s$ in these neighbourhoods with directions "similar" to the one defining $\gamma$ will have also conjugate points $q_s$. These conjugate points are "close" to $q$. These connecting geodesics conform a congruence containing $\gamma$ 
the set of conjugate points $q(s)$  becomes a caustic hypersurface. This result is independent on the underlying gravity model or the matter content.

The results presented may have applications related to the refractive index $n(u, \omega)$. In obtaining its upper plane singularity structure
the authors \cite{hollowood1}-\cite{hollowood2} employ the Penrose limit of the space time metric $g$ around the geodesic $\gamma$ under study. The resulting geometry is a gravitational wave. The profiles these references
consider posses caustics, but the authors wonder if the existence of caustics is a generic situation or if they may disappear when lifting the geodesics to full geometry.  In fact, geodesics curves in the approximated geometry 
may not correspond to genuine geodesics in the full geometry and therefore, the caustics may be an artificial effect of the Penrose limit. The proposition presented here shows that these caustics are a generic feature, and the proof
is given by formal homotopy arguments. Thus, given a full geometry and a geodesic with two conjugate points, as there are caustics around these points, the Penrose approximation employed in those references describe correctly from the qualitative
point of view
the physics corresponding to the full geometry. The approximated geodesics do indeed lift to a full geodesic and the approximated caustics do lift to a full caustic. The results about the singularities based on this approximation would not be spoiled by going to the full geometry.

In addition, the results of the proposition presented here implies that if the refractive index is interpreted as a function $n(u, \omega, \Lambda)$ with $\Lambda$ an appropriate parameter space denoting a congruence of geodesics, 
then the location of the singularities in the complex upper plane will be a function $\omega(\Lambda)$, which is continuous at least for an open containing $\Lambda_0$, being $\Lambda_0$ the parameter corresponding to the geodesic $\gamma$ under study.

A further interesting feature is that, for superluminal geodesics, which is a possibility that the authors  \cite{hollowood1}-\cite{hollowood2} consider, the set of points conjugate to $p$ may be continuous. The discreteness of the conjugate points
$q_n$ to $p$ for causal curves follows from the non degeneracy of the space time metric $g_{\mu\nu}$ when evaluated at the space $\mathbb{J}[u]$ of non trivial Jacobi fields at the parameter value $u$. This is exemplified in the lemma 2 of the previous section.
The discreteness follows from the fact that this non degeneracy implies that $\mathbb{J}[u]\oplus \mathbb{J}[u]^\perp=\mathbb{R}^n$. However, for a congruence of spatial geodesics $\gamma_s$ the Jacobi fields parallel to its tangent vector $u^\alpha$ are never vanishing, which means that
the space $\mathbb{J}[u]$ contains this spatial direction deleted. The resulting metric $g^\perp$ to perpendicular to this direction is still semi-riemannian and contains space like and time like directions. This implies that there  may exist Jacobi fields $v^\alpha$ in $\mathbb{J}[u]$ for which $g(v, v)=0$.
These vectors are parallel and perpendicular to themselves, and thus in this situation it is not warranted that $\mathbb{J}[u]\oplus \mathbb{J}[u]^\perp=\mathbb{R}^n$. This relation was crucial for the proof of Lemma 2 about discreteness and, if it is spoiled, discreteness is not assured. 

The possibility of a non discrete set of conjugate points to $p$ along  may result in the presence of a continuum of images of a given object, and was pointed out already in \cite{piccione1}. This continuous set of images are only possible in the non causal setting and may be employed as a tool for distinguishing superluminal effects. In addition, evaporation of conjugate points may be possible when varying along spatial geodesics in a given congruence as well since, as shown along the text, the identification between the Maslov index, the focal index and the geometrical one may fail if the metric $g^\perp$ is degenerate. Thus, in this situation, the Maslov index may not properly be counting the number of conjugate points along a geodesic, and the homotopy arguments just presented do not prevent the caustics to evaporate.

It should be mentioned that the consequences of superluminality are not necessarily restricted to geodesics. In fact, the birrefrigent curves considered in \cite{shorecaus}-\cite{shoreaus3}
are not geodesics from the strict point of view, as they are geodesic with respect to a deformed metric $G_{\mu\nu}$ not equal to $g_{\mu\nu}$. 
It may be interesting to construct a congruence of such curves and to understand if they intersect,
thus giving to an adapted definition of the notion of a conjugate point. This may be achieved by finding the analogous of the Jacobi equation in this context.
Furthermore, it may be interesting to see if the presence of caustic surface is generic in this context or not. We  have played 
with these birrefrigent curves and possible Jacobi field analogous, but at the moment we were unable to find a conclusive result. The problem we faced is that we were unable to find a suitable conserved symplectic form $\omega$
such as the one employed in the text for constructing lagrangian planes. When this obstacle appeared, we lost the analogy with the methods presented here. However, we hope to overcome these difficulties in a future.

Not less important, it should be mentioned that superluminality is not restricted to "light", and there are several works that study the velocity of the graviton, and there is no conclusive evidence yet that it is given by $c$. 
Several consequences of a superluminal graviton are discussed in detail in \cite{caus9}-\cite{caus12}, where causality bounds are presented. These references however are considering low energy effects coming from integrating
massive modes in gravitational theories, and their discussion is mainly restricted to low frequencies. Nevertheless the studies of generic conjugate points in non geodesic spatial curves may be relevant in this context as well.
We leave this for a future publication.

\section*{Acknowledgments}
O.P. S is supported by CONICET, Argentina and by the Grant PICT 2020-02181.

\appendix 
\section{A  brief mathematical interpretation of the quantities $Q_{ij}$}

The following discussion is very succinct, and the reader is referred to more mathematical references for further details.

As discussed along the text, the solutions of the Jacobi equation give rise to a lagrangian subspace $L_1(u)$ of $P$. At a conjugate point with parameter $u_1$ there exists a Jacobi field such that $v^a(u_1)=0$, and this means that $L_1(u_0)\cap L_1(u_1)\neq \{0\}$. Thus, it is of interest to study intersection theory
in the space of lagrangian subspaces of the phase space $P$. This space is known as the Lagrangian Grassmanian $G(n, P)$. 

Every point in the Grassmanian $G(n, P)$ represent a lagrangian subspace $L$, and it follows that $L_1(u)$ represents a path in $G(n, P)$. By taking into account that the real quantities $C_{ij}$ in (\ref{mitre}) are symmetric, it is deduced that the Dim $G(n, P)=n(n+1)/2$. It should be emphasized that this space is not oriented, this feature may depend on the dimensions \cite{fuks}.
In any case, the intersection of a given lagrangian space $L$ with other lagrangians in $P$ induce a hierarchy of points in $G(n, P)$ given by
$$
\Lambda^k(L)=\{l\in G(n,P)|\;\text{Dim} (l\cap L)=k\}.
$$
By the definition  of $\Lambda^k(L)$ given above, it is obvious that
$$
G(n, P)=\cup_{i=0}^n \Lambda^k(L).
$$
On the other hand, the set $\Lambda^0(L)$ represents the lagrangian subspaces complementary to $L$ and it is of no interest in the study of conjugate points. Thus, given the initial set $L_1(u_0)$, which will be denoted below as simply $L_0$,
the interest is focused on the Maslov set
\be\lb{maslovset}
M(L_0)=G(n,P)-\Lambda^0(L_0)=\cup_{i=1}^n \Lambda^k(L_0),
\ee
as this set represent the lagrangian subspaces $l$ whose intersection with $L_0$ is non trivial. These non trivial  intersections are the ones that signals the presence of conjugate points.

The set $\Lambda^k(L)$ corresponds to matrices $C_{ij}$ in (\ref{mitre}) which $k$ zero eigenvalues. This space has codimension $k(k+1)/2$ in $G(n, P)$. The regular set $\Lambda^1(L)$ has codimension one
and $\Lambda^2(L)$ has codimension three in $G(n, P)$. The set $M(L_0)$ itself is a cycle of codimension one.

Pretend for some moment that orientability for $G(n, P)$ holds. As the sympletic form $\omega$ defined in section 2 is non degenerate, if a lagrangian plane decomposition $P=L_1\oplus L_2$
has been found then, given a vector $v\in L_2$, the dual vector
$$
v^\ast=\omega(v,\cdot),
$$
may be constructed. Thus the conserved form $\omega$ induces the identification $L_1^\ast \simeq L_2$ of the dual space of $L_1$.
In addition, given any vector $v_2$ in $L_2$, one has that $$\omega(v_2+v_1, \cdot)|_{L_1}=\omega(v_2, \cdot)|_{L_1},$$ if $v_1$ belongs to $L_1$. This induces the following identification $L_1^\ast\sim P/L_1$. The space of lagrangian subspaces complementary to a given one $L_1$  is described in (\ref{mitre}) in terms of the symmetric $n\times n$ matrices
$C_{ij}$. By identifying $C_{ij}$ with $C=C_{ij} e_i\otimes e_j$ it is seen that this space can be identified with the quadratic forms $L_1\otimes L_1$. As $L_1\simeq L_2^\ast$ it is seen that
the $L_1\otimes L_1$ can be identified further with   $L_2^\ast \otimes L^\ast_2$. 

On the other hand, the tangent space $T_A(G(n, P))$, which is fundamental in the definition of the Maslov index, can be characterized as follows. Note that the formula
(\ref{lijas}) given by
$$
F_i(u)=F_i(u_f)+L_{ij}(u_f, u) E_j(u_f),
$$
shows that for $u=u_f+\delta u$ the quantities $L_{ij}(u_f, u)$ are approximately describing the directions in the grassmanian $G(n, P)$ emanating from the "point" $L_2(u_f)=\text{span}\{F_i(u_f)\}$.
A tangent vector in the tangent space $T_{L}G(n,P)$ is parameterized by $Q_{ij}=\dot{L}_{ij}$. In other words, $Q_{ij}$ represent the directions along a given point in the tangent space of the Grassmanian. 

The cycle $\Lambda^1(L_0)$ in the definition (\ref{maslovset}) can be interpreted as an element of $H^1(\Lambda, Z)$, and represents the lagrangian paths with regular intersection with $L_0$.
It is known as the Arnold-Maslov cycle \cite{arnold1}-\cite{arnold2}. The fact that is a codimension one cycle implies that, if  it had an orientation, this will be defined by elements $A$ of $\Lambda^1(L)$
by taking $\dot{A}$ of $T_A(G(n, P))$ which are transverse to $A$. In this context, transversality means the following. The results of the previous paragraphs shows that there is an identification
between $\dot{A}$ and a quadratic form in $A^\ast \otimes A^\ast$, which does not restrict to zero in $A$. If it was possible to give an orientation, then this will be defined
by the transverse vectors on which the quadratic form is positive definite, together with those on which is negative definite.
In these terms, given  a curve $\gamma:[a, b]\to \Lambda$ with endpoints not in $\Lambda^1(L_0)$, the intersection number with the Maslov cycle $\gamma\cdot \Lambda^1(L_0)$
may be defined, and enlarged to singular intersections by use of intersection theory. It is tempting to identify this intersection number it with the Maslov index. This picture however is not so straightforward. The problem is that, in general, the mentioned orientability is not warranted, and is dimension dependent. Thus, these definition have to be properly
modified to include these more complicated situations. The reader is referred to the mathematical works \cite{helfer2}, \cite{piccionelibro} for a large amount of details about this features. The tangent space parametrized by $Q_{ij}$
plays a fundamental role in the full definition of the Maslov index.

Despite these problems, the homotopy properties of the Maslov index may be characterized in terms of a formalism known as lagrangian planes \cite{robbin1}, \cite{mcduff}, which is helpful for our purposes. This formalism will be described below in certain detail, as is the one employed along the text.

\section{A more formal definition of the Maslov index}

As remarked in the previous sections, the quantity of real interest in the study of causal geodesics is the geometrical index $i_{geom}$ defined in (\ref{igeom}).
This quantity was identified the focal index $i_{focal}$ defined in (\ref{ifocal}) and, after some work, with the Maslov index (\ref{Maslov}). Af first sight, this may be a convoluted path for dealing with the problem, as the quantities are equivalent but their definition becomes more involved. The point for doing that is that the Maslov index is well studied in the literature \cite{robbin1}-\cite{leray}. In particular, some of its properties under path deformation are well understood \cite{robbin1}. The intention of the present work is to understand the behavior of conjugate points under geodesic perturbations, and for this reason this identification may be useful.

Still, in order to make a better comparison with literature, the formalism of lagrangian frames will be introduced. 
By expressing the Maslov index in this formalism, and by showing the equivalence with (\ref{Maslov}), some conclusions about homotopic deformations will be drawn immediately.

\subsection{The formalism of lagrangian frames}


Consider again some decomposition $P=L_1\oplus L_2$ of  some even dimensional space $P$, equipped with a sympletic 2-form $\omega$, which is expressed as the sum of two complementary lagrangian planes $L_1$ and $L_2$ with basis $e_a$ and $f_a$ respectively, see (\ref{ls}). For studying geodesics, $P$ may be identified with the phase space described in the previous sections. However, the following discussion is more general, not restricted solely to the geodesics case. In the basis $e_a$ for $L_1$ and $f_a$
for $L_2$ the following expression $\omega=\delta_{ab}e^a \wedge f^b$ is obtained.
Assign to the vector field $e_1$ a the $n$-components row vector $u^{T1}=(1,0,..,0)$, to $e_2$ the vector $u^{T2}=(0,1,0,...,0)$, and so on until $e_n$ is identified with the vector $u^{Tn}=(0,..,1)$. Then the space of lagrangian planes may be then characterized by the linear transformations $Z: R^n\to R^{2n}$ of the form
$$
Z=\left[
\begin{array}{c}
A \\
B
\end{array}\right],
$$
with $A$ and $B$ being $n\times n$ matrices, as follows. Given any of such transformations $Z$, the following $n$-dimensional subspace $L$ of $P$ 
$$
L=\text{span}\{Z u^1,..., Zu^n\},
$$
may be constructed. Explicitly, this space corresponds to the generators $L=\text{span}\{g_1,..., g_n\}$ given by
\be\lb{bazar}
g_i=A_{ji} e_j+B_{ji} f_j.
\ee
The matrices $A$ and $B$ are at the moment arbitrary.  But if $L$ is requested to be a lagrangian plane, that is, 
$\omega(g_i, g_j)=0$ for every choice of the indices $i$ and $j$, then the relations $\omega(e_i, e_j)=\omega(f_i, f_j)=0$ and $\omega(e_i, f_j)=\delta_{ij}$
imply that
\be\lb{compaty}
A^TB=B^TA.
\ee
A matrix $Z$ with components $A$ and $B$ satisfying the relation (\ref{compaty}) is known as a lagrangian frame for $L$. By varying along all the possible matrices $Z$ satisfying this condition the space of lagrangian planes is covered. 

%

\subsection{The Maslov index in the formalism of lagrangian frames}
Equipped with the formalism of lagrangian frames, a new definition of the Maslov index may be found.
Consider, as usual, the Jacobi problem (\ref{jaco}) with the initial conditions (\ref{init}). As explained in the previous sections, this give rise to a lagrangian path $L_1(u)$ in $G(n,P)$. The phase space of the problem may be decomposed as $P=L_1(u_0)\oplus L_2(u)$, the space $L_2(u)$ is any complementary lagrangian to $L_1(u)$.
A fundamental quantity for defining the Maslov index is the tensor $Q=\dot{L}$ found in (\ref{qform}).
This quantity, as seen from (\ref{derivv}), can be expressed as 
\be\lb{cu}
Q(v)=\frac{d\omega(v, v+w(u))}{du}\bigg|_{u=0}.
\ee
with $v \in L_1(0)$ and $w\in W(u)$ such that $v+w(u)\in L_1(u)$. Here $W$ is any lagrangian complementary to $L_1(u)$, not necessarily $L_2(u)$. This definition in fact should be independent on the choice of $W$.  

Note that the initial parameter in the definition of $Q(v)$ has been set $u_0=0$, there is no real loss of generality with this choice.

At the point $u=0$, fix a basis $f_i$ for $L_1(0)$ and a basis $e_i(0)$ for some complementary lagragian $W$, not necessarily equal to $L_2(0)$. Then the frame $Z(u)$ for $L_1(u)$ generically given by
$$
Z_1(u)=
\left[
\begin{array}{c}
A(u) \\
D(u)
\end{array}\right],
$$
is constrained by the initial condition
$$
 Z_1(0)=
\left[
\begin{array}{c}
0 \\
I
\end{array}\right].
$$
In other words, for $u=0$ the basis for $L_1(u)$ should reduce to $f_a$.
Thus,  for small $u$ the lagrangian frame  $Z(u)$ is 
\be\lb{chus}
Z_1(u)=
\left[
\begin{array}{c}
A(u) \\
I+\delta D(u)
\end{array}\right],
\ee
with $\delta D(u)$ and $A(u)$ matrices composed by infinitesimal quantities, and such that $\delta D(0)=A(0)=0_{n\times n}$. The condition for $Z(u)$ to be a lagrangian frame, that is, that $D^T A=A^T D$,
implies that $A^T(u)\simeq A(u)$ for $u$ small enough. This means that for small $u$ the matrix $A$ is approximately symmetric.
Therefore
\be\lb{lt}
L_1(u)\simeq\text{span}\{f_i+A_{ji}(u)e_j\},
\ee
for $u$ small enough. 

On the other hand, a lagrangian  space complementary to $L_1(u)$ is described generically 
at $u=0$ as
$$
W=\text{span}\{e_i+C_{ji}f_j\},
$$
as shown above in formula (\ref{mitre}). Thus, the lagrangian $L_2(u)$ is expected to be described for any $u$ by a lagrangian frame of the form
$$
Z_2(u)=
\left[
\begin{array}{c}
I+B(u) \\
C(u)
\end{array}\right],
$$
with $B_{ij}(0)=0$ and $C_{ij}(0)=C_{ij}$. Therefore, given a vector $v+w(u)$ with $v\in L_0$, $w(t)\in W$  one has for $u$ small that
$$
v=v^a e_a,\qquad w(u)=w^a(u) (f_a+C_{ba}(u)e_b),
$$
where the effect of $B(u)$ is neglected for small parameter values.
In these terms
$$
v+w(u)=(v^a+ C_{ab}w^b)e_a+w^a f_a.
$$
However, $u^a$ and $w^a$ are not independent since, as stated above, $v+w(u)\in L_1(u)$. This, together with description (\ref{lt}) of $L_1(u)$ implies that
$$
w^a=A_{ab}(v^b+C_{bc}w^c).
$$
With the help of the last formula, it is found then that
$$
\omega(v, v+w(u))=\omega(v, w(u))=\delta_{ab}v^a w^b(u)=\delta_{ab}A_{ad}(v^d+C_{cd}w^c) v^b.
$$
By taking this and the fact that $A_{ab}(0)=0$ into account, and by introducing the notation \be\lb{escalar}<x, y>=\delta_{ab}x^a y^b,\ee it is easily calculated that
\be\lb{rojo}
Q(v)=\frac{d\omega(v, v+w(u))}{du}\bigg|_{u=0}=<v, \dot{A}(0) v>.
\ee
This formula  reproduces the third formula of the proof of Theorem 1.1 in the reference \cite{salamon}.
The quantity $Q(v)$ in (\ref{rojo}) does not depend on $C_{ij}$ and thus is the same no matter the choice of the complementary lagrangian.
This is an important check.

Another link with \cite{salamon} comes by employing a generic lagrangian frame for  $L_1(u)$, not necessarily equal to (\ref{chus}).
As the definition of $Q(v)$ is independent on the choice of the complementary lagrangian, one may choose $L_2(u)$ as the one described by the frame
$$
W_2(u)=
\left[
\begin{array}{c}
U(u) \\
0
\end{array}\right],
$$
with $U(0)=I$. For the lagrangian $L_1(u)$ chose the frame
$$
W_1(u)=
\left[
\begin{array}{c}
Y(u) \\
X(u)
\end{array}\right],
$$
with the component matrices not constrained by any particular condition except the lagrangian frame one $X^T Y=Y^T X$.
This means that $L_1(u)=\text{span}\{g_i\}$ with
\be\lb{gi}
g_i=Y_{ji}(u) e_j+X_{ji}(u) f_j.
\ee
The initial vector $v$ and the variation $w(u)$ for small $u$ are then given by
$$
v=V^ig_i,\qquad w(u)=w^i U_{ji}(u)e_j.
$$
The vector $V^i$ are the components of $v$ in the basis $g_i$. Its projection on the basis $e_i$ and $f_j$ are
given by
$$
v^i=V^j Y_{ij}(0), \qquad v^j=V^i X_{ji}(0),
$$
respectively. The vector $V^i$ is defined  in terms of $v^i$
by these relations. 
By denoting $w^i(u)=w^j U_{ij}(u)$, it follows that
$$
v+w(u)=(V^i Y_{ji}(0)+w^j(u))e_j+V^i X_{ji}(0) f_j.
$$
There is a further relation between the above defined quantities, that come from the fact that $v+w\in L_1(u)$. 
By employing the notation $(v+w)^l$ for the components of $v+w$ in the basis $g_l$, it is clear that
$$
(V^i Y_{ji}(0)+w^j)e_j+V^i X_{ji}(0) f_j=(v+w)^lg_l.
$$
By use explicit of (\ref{gi}) it is found that
$$
 Y_{ji}(0)V^i+w^j= Y_{jl}(u)(v+w)^l,\qquad  X_{ji}(0)V^i= X_{jl}(u)(v+w)^l.
$$
By then eliminating the quantity $(v+w)^l$ in the last expressions, it is found that
\be\lb{ruls}
Y_{ji}(0) V^i+w^j=Y_{jl}(X^{-1})^{kl} X_{ki}(0)V^i
\ee
In these terms, by using the notation (\ref{escalar}) and by taking into account that $\omega(e_i, f_j)=\delta_{ij}$, it is found that
$$
\omega(v, w(u))=<X(0)V, w(u)>.
$$
From here it follows that
$$
Q(v)=\frac{d\omega(v, v+w(u))}{du}\bigg|_{u=0}=<X(0)V,\dot{w}(0)>.
$$
The last formula can be further worked out from (\ref{ruls}) as
\be\lb{lab}
Q(v)=<X(0)V,\dot{Y}(0)V>-<X(0)V,Y(0)X^{-1}(0)\dot{X}(0)V>.
\ee
The second term can be worked further. A simple calculation shows that
$$
<X(0)V,Y(0)X^{-1}(0)\dot{X}(0)V>=\delta_{ij} X_{jl} V^l Y_{jk} (X^{-1})_{km} \dot{X}_{mn} V_n.
$$
The condition (\ref{compaty}) for lagrangian frames  $X^TY=Y^TX$, that is, 
$X_{jl}Y_{jk}=Y_{jl}X_{jk}$ can be employed for cancelling out the terms proportional to $X_{ij}$ and $(X^{-1})_{ij}$ in the last formula. In these terms,  it is found
that
$$
<X(0)V,Y(0)X^{-1}(0)\dot{X}(0)V>=<\dot{X}(0)V,Y(0)V>,
$$
and the quantity (\ref{lab}) becomes
\be\lb{lab2}
Q(v)=<X(0)V,\dot{Y}(0)V>-<\dot{X}(0)V,Y(0)V>.
\ee
This formula reproduce the second point of the statement of the Theorem 1.1 of reference \cite{salamon}, up to some minor change of notation. The theory developed
in that classic reference lead to the homotopy properties stated along the text.

\end{document}